\documentclass[11pt,twoside]{article}
\usepackage[utf8]{inputenc}
\usepackage[T1]{fontenc}
\usepackage{amsmath}
\usepackage{amsfonts}
\usepackage{amssymb}
\usepackage{xcolor}
\usepackage{mathtools}
\usepackage{hyperref}
\usepackage{doi}
\usepackage{array}
\newcolumntype{s}{>{\footnotesize}l}
\usepackage{booktabs}
\hypersetup{pdfborder={0 0 0},
	colorlinks=true,
	linkcolor=black,
	citecolor=black,
	urlcolor=black}
\usepackage{cleveref}
\usepackage{fourier}

\date{Compiled \today}

\newcommand{\pkg}{\textbf}

\newcommand{\code}{\texttt}
\newcommand{\soft}{\textsf}
\usepackage[top=3cm, bottom=3cm, left=2.25cm, right=2.25cm]{geometry}
\usepackage[calcwidth,  sf,  big, compact]{titlesec}
\titleformat{\section}
{\normalfont\scshape \Large}{\thesection}{1em}{}
\usepackage{fancyhdr}
\usepackage{setspace}
\usepackage[authoryear]{natbib}
\usepackage{cleveref}

\titleformat{\section}[block]{\large \bf }
{  {\thesection.}}{4pt}{   }
\titleformat{\subsection}[block]{\itshape}
{  {\thesubsection.}}{4pt}{   }

\title{A modeler's guide to extreme value software}
\author{L\'eo R. Belzile\thanks{HEC Montréal, Department of Decision Sciences, HEC Montréal (\texttt{leo.belzile@hec.ca})}, Christophe Dutang\thanks{CEREMADE, CNRS and Université Paris Dauphine - PSL, Paris, France}, Paul J. Northrop\thanks{Department of Statistical Science,
University College London, United Kingdom}, and  Thomas Opitz\thanks{Biostatistics and Spatial Processes, INRAE, Avignon, France}
}
\date{}

\begin{document}
\maketitle
\abstract{
This review paper surveys recent development in software implementations for extreme value analyses since the publication of \cite{stephensongilleland06} and \cite{Gilleland/Ribatet/Stephenson:2013},  here with a focus on numerical challenges. We provide a comparative review by topic and highlight differences in existing routines, along with listing areas where software development is lacking. The online supplement contains two vignettes
providing a comparison of implementations of frequentist and Bayesian
estimation of univariate extreme value models.
}

\section{Introduction}
 Extreme value analysis has seen strong development over the years. While software development typically lags behind methodological developments due in part to lack of recognition of the effort needed to provide reliable software, reproducibility requirements and individual efforts have led to a growth in the coverage of statistical methods. Many procedures developed in the last decades are now 
  available, but the diversity of numerical implementations complicates somewhat the choice of routine to adopt. 

Our intention, rather than to solely provide a catalog of existing software, is to discuss and compare existing implementations of statistical methods and to highlight numerical issues that are of practical importance yet are not typically discussed in theoretical or methodological papers. Our work also provides an update including the most recent software development since the reviews of \citet{stephensongilleland06,Gilleland/Ribatet/Stephenson:2013,Gilleland2016}.

Given its ongoing popularity, we focus on implementations using the \soft{R} programming language, unless stated otherwise. The \href{https://cran.r-project.org/}{Comprehensive R Archive Network (CRAN)} \href{https://cran.r-project.org/web/views/ExtremeValue.html}{Task View on Extreme values} provides an extensive list of package functionalities organized by topics; we follow this approach and broadly separate implementations into univariate, multivariate and functional extremes rather than present functionalities package by package. Using the \pkg{RWsearch} package \citep{R-RWsearch}, we automated the process of searching for extreme-related packages on the CRAN and inspected all of the packages that have ``extreme value'' or ``peak over threshold'' as keywords in the package description. Additional searches were done for unpublished packages.

\section{Univariate extremes}
\subsection{Asymptotic theory for univariate extremes}

The starting point for univariate extreme value analysis is the extremal types theorem.  For convenience, we state the result in terms of point process representations, i.e., of the behavior of random point clouds: let $X_i$ be independent and identically distributed random variables with distribution $F$, where $F$ has lower endpoint $x_*= \mathrm{inf}\{x: F(x) > 0\}$ and upper endpoint $x^* = \sup\{x: F(x) < 1\}$. If there exist normalizing sequences $a_n>0$ and $b_n \in \mathbb{R}$ such that the distribution of the bidimensional point process
\[P_n =\left\{ \left(\frac{i}{n+1}\right), \frac{X_i-b_n}{a_n}, i = 1, \ldots, n\right\}\]
converges to an inhomogeneous Poisson point process on sets of the form $(a, b) \times (z, \infty)$ for $0  \leq a \leq b \leq 1$ and  $z>z_*=\lim_{n \to \infty} \{(x_*-b_n)/a_n\}$, the intensity
measure of the limiting process,  which gives the expected number of points falling in a set,  is \citep[cf.,][]{Coles:2001}
\begin{align}
\Lambda\{(a, b) \times (z, \infty)\} = \begin{cases}
(b-a)\left(1+ \xi \frac{z-\mu}{\sigma}\right)_{+}^{-1/\xi}, &\xi \neq 0\\
(b-a)\exp\left(\frac{z-\mu}{\sigma}\right), &\xi = 0
\end{cases}, \qquad \mu \in \mathbb{R}, \sigma > 0, \label{eq:pp_conv}
\end{align}
 where $x_{+} = \max\{0, x\}$.
Equipped with this convergence statement, we can consider various tail events. For example, the limiting distribution of the maximum of $n$ observations is
\begin{align}
 \lim_{n \to \infty} \Pr\left(\frac{\max_{i=1}^n X_i - b_n}{a_n} \leq x \right)  =\exp\left[-\Lambda\{(0,1) \times (x, \infty)\}\right]. \label{eq:gevconv}
\end{align}
The right hand side of \Cref{eq:gevconv} is the distribution function of the generalized extreme value distribution, say $G(x)$, with location parameter $\mu \in \mathbb{R}$, scale parameter $\sigma \in \mathbb{R}_{+}$ and shape parameter $\xi \in \mathbb{R}$,

defined on $\{x \in \mathbb{R}: \xi(x-\mu)/\sigma > -1\}$. For historical reasons, the distribution is categorized based on the sign of $\xi$ in so-called ``domains of attraction''. If $\xi<0$, the distribution has a bounded upper tail, $\xi=0$ leads to an exponential ``light'' tail and $\xi>0$ to a ``heavy tail'' with polynomial decay with finite moments only of order $r < 1/\xi$.

We can also consider conditional exceedances of a threshold $u$ as the latter increases to the upper endpoint $x^*$; for threshold exceedances over $u > z_*$, there exists $a_u>0$ such that
\begin{align}
\lim_{u \to x^*}\frac{\Pr(a_u^{-1}X  > x+u)}{\Pr(a_u^{-1}X > u)} & = \frac{\Lambda\{(0, 1) \times (x+u, \infty)\}}{\Lambda\{(0, 1) \times(u, \infty)\}} = 1-H(x),\label{eq:gpconv}
 \intertext{where}
 1-H(x) &=\begin{cases}
 \left\{1+\xi \left(\frac{x}{\sigma_u}\right)\right\}_{+}^{-1/\xi}, & \xi\neq 0,\\
 \exp(-x/\sigma_u) & \xi = 0;
\end{cases}\label{eq:gpdist}
\end{align}
and $\sigma_u = \sigma + \xi(u-\mu)$. The right-hand side of \Cref{eq:gpdist} is the survival function of the generalized Pareto distribution with scale $\sigma_u$ and shape $\xi \in \mathbb{R}$.

\subsection{Maximum likelihood estimation}

We can approximate the log likelihood by taking the limiting relations of \Cref{eq:pp_conv,eq:gevconv,eq:gpconv} as exact for the maximum of a finite block of $m$ observations or for exceedances above a large quantile $u$; the unknown normalizing constants $a_n$, $b_n$ (respectively $a_u$) are absorbed by the location and scale parameters. For example, the log likelihood obtained through the limiting inhomogeneous Poisson point process for a sample of $N$ independent observations with exceedances $y_1, \ldots, y_{n_u}$ is
\begin{align}
\ell(\mu, \sigma, \xi; \boldsymbol{y}) &=  - n_u \log(c\sigma)  - \sum_{i=1}^{n_u} \left(1+\frac{1}{\xi}\right)\log\left\{1+\xi\left( \frac{y_i-\mu}{\sigma}\right)\right\}_{+} \nonumber\\& \qquad - c \left\{1+ \xi \left( \frac{u-\mu}{\sigma}\right)\right\}^{-1/\xi}_{+}, \qquad \mu, \xi \in \mathbb{R}, \sigma >0. \label{eq:ppp_lik}
\end{align}
The quantity $c$, which does not appear in \Cref{eq:pp_conv}, is a tuning parameter \citep[][\S~7.5]{Coles:2001}: if we take $c=N/m$, the parameters of the point process likelihood correspond to those of the generalized extreme value distribution fitted to the maximum of blocks of $m$ observations. This parametrization however induces strong correlation between the parameters $(\mu, \sigma, \xi)$.

\begin{figure}
 \centering
 \includegraphics[width = 0.9\textwidth]{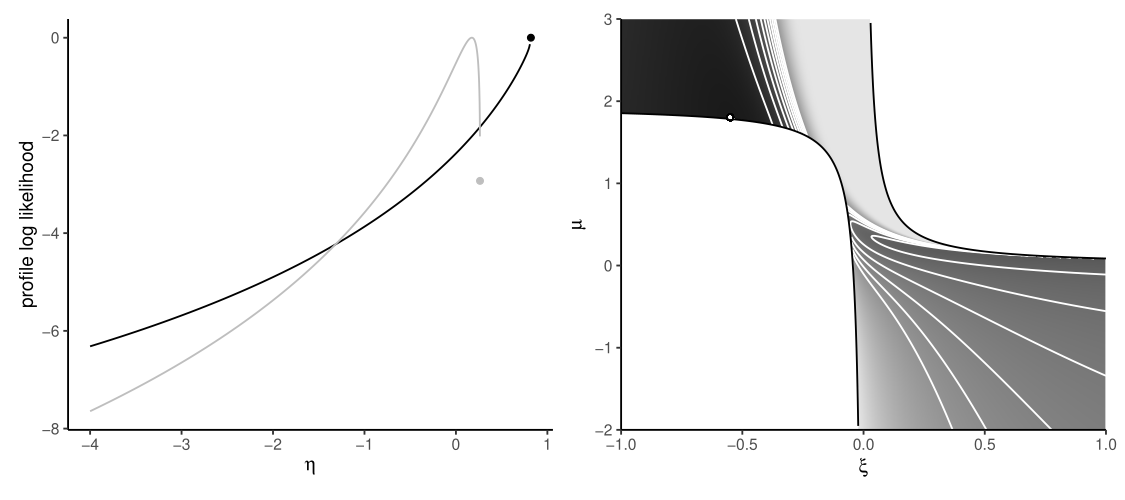}
 \caption{Left: profile log likelihood of $\eta=-\xi/\sigma$ for a generalized Pareto distribution with scale $\sigma$ and shape $\xi$. The black line shows one data set for which the conditional maximum $\widehat{\xi}_{\widehat{\eta}}=-1$ lies on the boundary of the parameter space (black) and one with $\widehat{\xi}_{\widehat{\eta}} > -1$ (grey).
 Right: conditional log likelihood surface for the inhomogeneous Poisson process at $\widehat{\sigma}$ for simulated data (larger values have darker grey-scale shade). The white dot indicates the maximum likelihood estimate, while the hyperbola defines the feasible region of the parameter space given by the support constraints. }
 \label{fig:ipp}
\end{figure}

Likelihood-based inference for extreme value distributions is in principle straightforward, even if there is no closed-form solution for the maximum likelihood estimators (\textsc{mle}).
The latter must solve the score equations $S(\boldsymbol{\theta})=\boldsymbol{0}$ (i.e., the log likelihood must have gradient zero) unless $\widehat{\xi}=-1$, but with nonlinear inequality constraints depending on the sign of $\xi$ (e.g., $\max(x) \leq \mu-\sigma/\xi$ if $\xi < 0$ for the generalized extreme value distribution). Constrained gradient-based optimization algorithms are thus logical choices for finding the \textsc{mle}. Many numerical implementations of the log likelihood simply return very large finite values for parameter combinations outside of the support, which can impact the convergence of gradient-based optimization routines: the user is invited to check convergence of whichever software is employed by evaluating the score function at the \textsc{mle} configuration to make sure it is indeed a root of the score vector $S(\boldsymbol{\theta})$. Even then, the solution returned may not be a global maximum: \Cref{fig:ipp} shows the conditional log likelihood surface for an inhomogeneous Poisson process model, obtained by fixing the scale. The feasible region is defined by a hyperbola and features two local maxima; depending on the starting value, a gradient algorithm would converge to different values.

Particular attention must be paid to numerical overflow when implementing the likelihood, score and information matrix of the generalized extreme value distribution, especially for terms of the form $\log(1+\xi x)$ when $\xi \to 0$ for the information and  cumulants. For example, the expected information matrix satisfies $I_{\xi\xi}=f(\xi)/\xi^{-4}$ \citep{Prescott/Walden:1980} and the limit as $\xi \to 0$ is well-defined, but this expression is numerically unstable when $\xi \approx 0$. High precision functions such as $\code{log1p}$ can be used to alleviate this somewhat, and interpolation of the cumulants when $\xi \approx 0$ is recommended.

The extreme value distributions do not satisfy regularity conditions because their endpoint depends on the parameter values. Maximum likelihood estimators for both the generalized extreme value and generalized Pareto distributions fitted to respectively block maxima and threshold exceedances are nevertheless asymptotically Gaussian and the models are regular whenever $\xi > -1/2$ \citep[][\S~3.4]{Smith:1985,Drees/deHaan/Ferreira:2004,Bucher/Segers:2017,deHaan/Ferreira:2006}.

Moments of some of the $k$th order derivatives of the log likelihood exist only if the shape $\xi > -1/k$. Thus, when $\xi \leq -1$, the \textsc{mle} does not solve the score equation. The likelihood functions for the inhomogeneous Poisson point process of exceedances and the $r$-largest observations, the generalized Pareto, the generalized extreme value, are unbounded if $\widehat{\xi}<-1$, as there exists a combination of parameters that lead to infinite log likelihood values. This means one should restrict the parameter space to $\{\xi: \xi \geq -1\}$ and check that the solution does not lie on the boundary of the parameter space: for the generalized extreme value distribution, the conditional maximum likelihood estimator for $\xi=-1$ is $\widehat{\mu}=\overline{y}$ and $\widehat{\sigma} = \max(y) - \overline{y}$, whereas for the generalized Pareto distribution, $\widehat{\sigma}=\max(y)$, for the likelihood of the $r$-largest order statistics, $\widehat{\sigma} = \{\max(y)-\overline{y}_{(r)}\}/r, \widehat{\mu}=\max(y) - \widehat{\sigma}$, etc.

The (lack of) existence of cumulants also impacts the calculation of standard errors, as elements of the Fisher information matrix are defined only if $\xi > -1/2$. Most software implementation compute standard errors based on the inverse Hessian matrix computed via finite-difference, but these are misleading if $\xi \in [-1/2,-1)$.

We can sometimes deploy dimension reduction strategies to facilitate numerical optimization. For the generalized Pareto distribution, \cite{Grimshaw:1993} uses a profile likelihood to reduce the problem to a one-dimensional optimization. This is arguably one of the safest maximum likelihood estimation procedures and the exponential sub-case, for which the profile likelihood is unbounded, can be easily handled separately. The left panel of \Cref{fig:ipp} shows profile log likelihoods for two simulated datasets, including one for which $\widehat{\xi}$ lies on the boundary of the parameter space.

 As noted before, the \textsc{mle} of the parameters of the inhomogeneous Poisson point process are hard to obtain because of the strong dependence between parameters: the optimization in packages such as \pkg{ismev} \citep{ismev,Coles:2001} or \pkg{evd} \citep{evd} often fails to converge, mostly because of poor starting values and existence of several local maxima.  \cite{Sharkey:2017} propose a reparametrization that ensures approximate orthogonality of the parameters, but the following trick can also help with this: if the estimated probability of exceedance is small, the Poisson approximation implies
\[c \left\{1+ \xi \left( \frac{u-\mu}{\sigma}\right)\right\}^{-1/\xi} \approx n_u. \]
 We can thus fit a generalized Pareto distribution to threshold exceedances, whose maximum likelihood estimates we denote $(\widehat{\sigma}_u, \widehat{\xi})$, and then use as starting values for the point-process optimization routine
 \begin{align*}
\mu_0 = u - \sigma_0\{(n_u/c)^{-\widehat{\xi}}-1\}/\widehat{\xi}, \qquad \sigma_0 = \widehat{\sigma}_u\times (n_u/c)^{\widehat{\xi}}, \qquad \xi_0 = \widehat{\xi}.
\end{align*}

\subsubsection{Case study}
We performed some sanity checks for various implementations of maximum likelihood estimation routines and parametric models. Specifically, we verified that density functions are non-negative and evaluate to zero outside of the domain of the distribution, and that distribution functions are non-decreasing and map to the unit interval.  Certain packages have incorrect implementations of density and distribution functions.

To assess the quality of the optimization routines for extreme value distribution, we simulated exceedances and block maxima from parametric distributions with varying tail behaviors. We compared the maximum likelihood estimates returned by default estimation procedures for different packages for simulated data, checking that the value returned is a global optimum and the gradient evaluated at the value is approximately zero whenever $\widehat{\xi} > -1$.  The purpose of the exercise was to check the reliability for a range of sample sizes; when systematic differences in maximum log likelihood and/or parameter estimates arose compared to other packages, then they are attributable to poor starting values, incorrect implementation of the density function, lack of handling of boundary constraints or the choice of optimization algorithm.

\begin{figure}
 \centering
 \includegraphics[width = 0.9\textwidth]{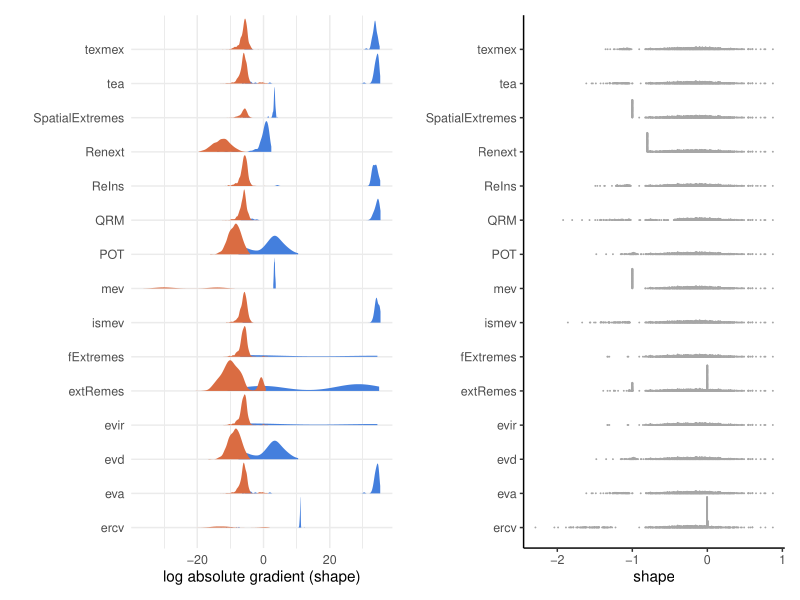}
 \caption{Diagnostics for maximum likelihood estimation of the generalized Pareto distribution based on 1000 samples simulated from a gamma distribution with shape 3 and scale 2 and a random number of exceedances (average 20) above the theoretical 0.95 quantile. Left: density plots of the absolute value of log gradient $\partial \ell / \partial \xi$ evaluated at the maximum likelihood estimator ($\widehat{\sigma}, \widehat{\xi}$), split by simulations yielding a boundary case ($\widehat{\xi} = -1$, blue) and regular case ($\widehat{\xi} > -1$, orange); the $y$-axis scale for each package is different to ease visualization. Right: dot plots of the shape parameter estimates. Results for samples for which the numerical routines failed to converge are omitted.}
 \label{fig:GPsimu}
\end{figure}

\Cref{fig:GPsimu} shows results for $1000$ simulations, each based on $n=400$ observations from a gamma distribution with shape 3 and scale 2; we set the threshold to the 0.95 theoretical quantile of the distribution, which leads to an average of 20 threshold exceedances. The left panel shows the distribution of the gradient of the log likelihood of the generalized Pareto distribution evaluated at the maximum likelihood estimate over all replicates for the shape parameter. Most instances of non-zero gradient are attributable to boundary cases, with $\widehat{\xi}=-1$ not accounted for. Other discrepancies are due to differences in numerical tolerance, but the differences in log likelihood relative to the maximum over all routines are negligible in most non-boundary cases investigated.

The right panel of \Cref{fig:GPsimu} shows the source of some oddities: the \pkg{QRM} package has unexpected small spread and a positive bias for estimation of $\xi$, different from other packages because it fails more often when $\xi$ is negative due to poor starting values. Both \pkg{ercv} and \pkg{extRemes} fit have noticeable point masses at $\xi=0$, suggesting something is amiss. Many packages do not enforce boundary constraints and may return $\widehat{\xi} < -1$ with near-infinite gradient estimates, responsible for the large gradients in the left panel of \Cref{fig:GPsimu}: \pkg{Renext} returns a hard-coded lower bound that is larger than $-1$, while \pkg{SpatialExtremes} and \pkg{mev} correctly return $-1$.

By contrast, the optimization routines for the generalized extreme value distribution yielded similar behavior and nearly all packages give identical results. However, we noticed that some packages fare poorly when location or scale parameters are orders of magnitude larger than scaled components: since the generalized extreme value distribution is a location-scale family, scaling the data before passing them to the routine and back-transforming after the optimization may solve such issues.

\subsection{Regression modeling}
\label{ssec:regression}

Most data encountered display various forms of nonstationarity, including trends, seasonality and covariate effects, which the extreme value distributions cannot capture without modification. One can thus consider regression models in which the parameters of the extreme value distributions are functions of covariates or vary smoothly in space or time. These parameters may be suitably transformed via a link function to ensure that the functions satisfy the usual range or positivity constraints. If we  assume independent observations, then

maximum likelihood estimates, standard errors, etc. are obtained as before by maximizing the log likelihood function, which is now a function of the regression coefficients and of other parameters arising in the nonstationary formulation of the extreme value distribution.  In models with a relatively large number of parameters, it becomes useful to  include an additive penalty term in the log likelihood: for example, generalized additive models for the parameters include smooth functions (\emph{smooths} in short) via basis function representations (e.g., $B$-splines), with a penalty that controls the wiggliness of the estimated predictor functions, which is typically evaluated using the second-order derivative of the basis functions.
The obvious difficulty for numerical maximization of the log likelihood is again the presence of support constraints, since there are now  potentially as many inequality constraints as there are observations. A general advice for models with covariates is that inputs should be centred and scaled to facilitate the optimization.

The \pkg{ismev}, \pkg{texmex}, \pkg{eva} and \pkg{extRemes} packages allow users to provide a model matrix (containing one covariate in each of its columns) for each parameter of the generalized Pareto and generalized extreme value distributions, thus enabling generalized linear modeling of the parameters, while the \pkg{evd} package only allows for linear modeling of the location parameter of the generalized extreme value distribution and bivariate counterparts; in both cases, no penalty terms are added to the log likelihood, while \pkg{texmex} allows for $L_1$ and $L_2$ penalties for the coefficients. The \pkg{lax} package supplements the functionality of these, and other, packages by providing robust sandwich estimation of parameter covariance matrix and log likelihood adjustment \citep{Chandler/Bate:2007} for their fitted model objects. The \pkg{GEVcdn} package uses a neural network to relate the parameters of the generalized extreme value distribution with covariates \citep{Cannon:2010}.

The scale parameter of the generalized Pareto distribution,  $\sigma_u$, varies with the threshold $u$: it is recommendable to pay special attention to the parametrization of the scale and shape functions with covariates to ensure that the threshold stability property, which is used for extrapolation, is not lost \citep{Eastoe:2009}.  It may be tempting to use directly the likelihood of \cref{eq:ppp_lik} instead \citep[see][]{NJR:2016}. \cite{Chavez/Davison:2005} use an orthogonal reparametrization $(\eta, \xi)$, where $\eta=\sigma(1+\xi)$ along with bootstrap routines for uncertainty quantification; their generalized additive modeling framework is available via \pkg{ismev}.

Penalty terms are often used in regression models with many coefficients to control the local variability of the estimated predictors, and they allow obtaining useful estimates for coefficients whose value would otherwise  not directly depend on any observations in the sample (e.g., in the case where the regression function is intended to capture spatial variability of extreme value parameters but some regions of space do not contain any observation locations). Many general packages implement generalized additive modeling with some support for extreme value distributions, including \pkg{VGAM}  \citep{Yee.Stephenson:2007}.  The recent \pkg{evgam} package \citep{Youngman:2020b},  dedicated to extreme value models, uses the methodology of \cite{Wood/Pya/Safken:2016} for general distributions to marginalize out the regression coefficients  using Laplace's method to obtain estimates of the hyperparameters (e.g., variance and autocorrelation of regression coefficients) controlling the penalty strength and shape --- these are estimated simultaneously with all of the other parameters through maximum likelihood.

The \pkg{evgam}  package builds on generic model building tools available in packages such as \pkg{mgcv} \citep{Wood:2017book} and provides state-of-the-art methodology tailored for extremes, including generalized additive models for extreme value distributions, quantile regression and in addition functionalities for obtaining return levels for the nonstationary models.

\begin{table}
\centering
\begin{tabular}{lllll}
\toprule
package & functions & type & link  & model \\
\midrule
\pkg{ismev} & \code{gpd.fit}, \code{gev.fit} & linear  & custom & \textsc{gev}, \textsc{gp}\\
\pkg{evd} & \code{fgev} &  linear   & linear & \textsc{gev} (loc)\\
\pkg{texmex} & \code{emv} & linear & log (scale) &  --- \\
\pkg{eva} & \code{gevrFit}, \code{gpdFit} & linear & custom &\textsc{gevr}, \textsc{gp} \\
\pkg{extRemes} & \code{fevd} & linear  & linear/log (scale) &  --- \\
\pkg{ismev} & \code{gamGPDfit} & GAM  &fixed &\textsc{gp}\\
\pkg{GEVcdn} & \code{gevcdn.fit} & neural network &fixed & \textsc{gev} \\
\pkg{VGAM} & \code{gev}, \code{gp} & GAM & custom & \textsc{gev}, \textsc{gp}, *\\

\pkg{evgam} & \code{fevgam} & GAM &fixed &\textsc{gevr}, \textsc{gp}, *\\
\bottomrule
\end{tabular}
\caption{Functionalities for generalized linear (linear) or generalized additive modelling (GAM) of the parameters of extreme value distributions. Model families supported include generalized extreme value distribution (\textsc{gev}), generalized Pareto (\textsc{gp}), $r$-largest extremes (\textsc{gevr}) and more general families or special cases of extreme value distributions, miscellaneous (*).}
\end{table}

There usually is a Bayesian interpretation of such quadratic penalty terms as corresponding to some multivariate Gaussian prior distribution on the regression coefficients. Fitting regression (or \emph{multilevel}) models is natural in the Bayesian setting, and many of the packages discussed in the next section have capabilities for fitting multilevel models.

\subsection{Bayesian modeling}
\label{sec:bayesian}

\subsubsection{Generalities}

In the Bayesian paradigm, the likelihood of the data vector $\boldsymbol{Y}$ is combined with prior distributions for the model parameters $\boldsymbol{\theta} = (\theta_1,\ldots,\theta_m)^\top \in \boldsymbol{\Theta}$, with prior density  $p(\boldsymbol{\theta})$; we use the generic notation $p(\ldots)$ for various conditional and unconditional densities. The distribution of the data given the parameter is obtained from the likelihood function $p(\boldsymbol{y} \mid \boldsymbol{\theta}) \propto \exp\{\ell(\boldsymbol{\theta}; \boldsymbol{y})\}$, where $\propto$ denotes proportionality.

The posterior distribution,
\begin{align}
 p(\boldsymbol{\theta} \mid \boldsymbol{y}) = \frac{p(\boldsymbol{y} \mid \boldsymbol{\theta})p(\boldsymbol{\theta})}{ \int p(\boldsymbol{y} \mid \boldsymbol{\theta})p(\boldsymbol{\theta})\mathrm{d}\boldsymbol{\theta}}, \label{eqbayes}
\end{align}
 is proportional as a function of $\boldsymbol{\theta}$ to the product of the likelihood and the priors in the numerator, but the integral appearing in the denominator of \Cref{eqbayes} is untractable in general.  In such cases, the posterior density  $p(\boldsymbol{\theta} \mid \boldsymbol{y})$ usually does not correspond to any well-known distribution family, and posterior inferences about the components of $\boldsymbol{\theta}$ further involve marginalizing out the other components. For instance, to obtain the posterior density $p(\theta_1\mid \boldsymbol y)$ of the first parameter in $\boldsymbol{\theta}$, we have to evaluate the $(m-1)$-dimensional integral $\int p(\boldsymbol{\theta} \mid \boldsymbol{y})\,\mathrm{d}(\theta_2,\ldots,\theta_m)$.    Most of the field of Bayesian statistics revolves around the creation of algorithms that circumvent the calculation of the normalizing constant (or else provide accurate numerical approximation of the latter) or that allow for marginalizing out all parameters except for one.

Rather than a point estimate $\widehat{\boldsymbol{\theta}}$ of the parameter vector, the target of inference is the whole posterior distribution. The majority of Bayesian estimation algorithms are simulation-based, and their typical output  is an (approximate) sample drawn from the posterior distribution  $p(\boldsymbol{\theta} \mid \boldsymbol{y})$, from which any functional of interest can be estimated by Monte Carlo methods. Of particular interest is the posterior predictive distribution, which is obtained by simulating new observations from the response model by forward-sampling from $p(\boldsymbol{y} \mid \boldsymbol{\theta}^{(b)})$  one new observation for each draw of $\boldsymbol{\theta}^{(b)}$ from the posterior.

In simple problems,

exact sampling algorithms can provide independent and identical samples from the posterior.
This is the exception rather than the norm; most of the time, users resort to Markov chain Monte Carlo (MCMC) algorithms for more complex settings: these algorithms admit the posterior distribution as the stationary distribution of a Markov chain with appropriately designed transition probabilities and provide auto-correlated samples from it.
Another popular solution is through Laplace approximation for regression models  when multivariate Gaussian priors are put on the vector of regression coefficients arising in the latent layer of the model, from which observations are conditionally independent; see the discussion in Section~\ref{ssec:regression}.
In this setting, Laplace approximations give fast deterministic approximation of high-dimensional integrals, which avoids resorting to simulation-based MCMC estimation. Laplace approximations are particularly accurate when they are applied twice in a certain nested way, which is known as the integrated nested Laplace approximation \citep[INLA,][]{Rue/Martino:2009}, implemented in the general \pkg{INLA} package offering extreme value functionality for generalized Pareto and generalized extreme value distributions.

Despite the computational overhead associated, the Bayesian paradigm has many benefits, including the capacity to incorporate physical constraints and expert opinion through the prior distributions \citep{Coles/Tawn:1996}. It is easier and more natural to define hierarchical structures for parameters to pool information. For multivariate and functional extremes, priors can be used for regularization purposes to pool information, for instance across time and space.

\subsubsection{Specificity of extremes}

Readers wishing to learn more about Bayesian modeling for extreme values are referred to the extensive overview in \cite{Stephenson:2016}. While Bayesian inference for extreme value models does not differ much from that of general models, additional care is required with prior specification. For example, in order to get a well-defined posterior distribution, improper reference priors such as the maximal data information (MDI) and Jeffreys priors for $\xi$ may need to be truncated \citep{Northrop/Attalides:2016} to result in proper  (i.e., integrable) posterior distributions. \cite{Martins/Stedinger:2000} proposed using a shifted Beta distribution for $\xi$ to constrain the support of the latter to $[-0.5,0.5]$. Other popular choices are vague normal priors for location, log-scale and shape parameters, or else penalized complexity priors \citep{Simpson:2017,Opitz2018}. To avoid issues related to the finite and parameter-dependent lower endpoint in the generalized extreme value distribution for $\xi>0$, the \pkg{INLA} package implements so-called \emph{blended generalized extreme value distribution} that replaces the bounded lower distribution tail with the unbounded one of a Gumbel distribution through a mixture representation \citep{Castro2021}.

Three packages, \pkg{evdbayes}, \pkg{extRemes} and \pkg{MCMC4Extremes}, provide MCMC algorithms  for extreme value distributions, which implement so-called  random walk Metropolis--Hastings steps. \pkg{MCMC4Extremes} \citep{MCMC4Extremes} is superseded by the implementation of \pkg{evdbayes} \citep{evdbayes} which is more efficient, thanks to default tuning of proposal standard deviations and more flexible choices of priors. The user guide of \pkg{evdbayes} gives more details.

The underlying implementation of the MCMC algorithm for the function \code{posterior} in \pkg{evdbayes}

allows for a linear trend in the location parameter and supports inclusion of positive point mass at zero for the shape parameter (Gumbel distribution for generalized extreme value distribution/exponential distribution for the generalized Pareto distributions) by leveraging an MCMC implementation using reversible jumps, following \cite{Stephenson:2004}. The gamma priors for quantile differences, used for expert prior elicitation, are provided.

Contrary to most implementations, \pkg{evdbayes} returns a list of posterior samples and relies on methods implemented in \pkg{coda} \citep{coda} for diagnostic, summary and plots. The \pkg{extRemes} package \citep{Gilleland:2016} also has functionalities for computing posterior summaries for univariate extremes through the \code{fevd} function, which allows users to specify their own priors and proposal distributions, but the sampling is notably slower than in other packages

and more cumbersome to set up, as the default values are not adequate in most cases. Linear modeling of the  parameters with covariates is also possible, and Bayes factors for comparisons between models are also supported even if the methods used to compute them are not recommended. Package \pkg{texmex} also includes maximum a posteriori estimation and simulation from the posterior for  extreme value distributions (with linear modeling of covariates) via the function \code{evm}, but only with normal priors. Behind the scenes, the \pkg{texmex}  implementation uses an independent Metropolis--Hastings step with multivariate Cauchy/normal proposals with location vector and scale matrix based on a normal approximation to the posterior, using maximum a posteriori estimates. This translates into smaller autocorrelation (and thus larger effective sample size) than other package implementation, and it is the fastest of all MCMC implementations.

The data-driven prior proposed by \cite{Zhang/Stephens:2009}, reputed to give better results than maximum likelihood, is implemented in \pkg{mev} and is the default method for Pareto-smoothed importance sampling  \citep{Vehtari:2017} from the \pkg{loo} package \citep{loopackage}. However, because it uses the data to construct the prior, performance benchmarks alleging superior performances are misleading because of double dipping.

The current state-of-the-art method for sampling from the posterior of univariate models in simple analyses without covariates is the \pkg{revdbayes} package, which relies on the ratio-of-uniforms method to generate independent samples from the posterior distribution of the models. Use of advanced techniques such as mode relocation, marginal Box--Cox transformations and rotation can drastically improve the efficiency of this accept-reject scheme and make it very competitive. The ratio-of-uniforms method generates independent draws, thus avoiding the need to monitor convergence to the stationary distribution of the Markov chain and removing tuning parameters. The package uses the \pkg{Rcpp} interface \citep{Rcpp} to speed up sampling, and the sampling is an order of magnitude faster than other implementations.

While the aforementioned packages are dedicated to extreme value distributions, other popular programming languages could be used even if they would require users to implement likelihood functions themselves. Notably, the \soft{Stan} programming language \citep{Rstan}

uses Hamiltonian Monte Carlo, a state-of-the-art MCMC method, for simulating samples from the posterior distribution. The latter can easily be combined with multilevel models, but requires implementation of bespoke code for likelihood and priors that are specific to extreme value analysis. The Hamiltonian Monte Carlo sampling algorithm leads to rejection due to boundary constraints and leads to incorrect posterior draws for, e.g., the generalized extreme value distribution when $\xi \approx 0$; this can be corrected by using a Taylor series approximation. The \href{https://www.mathworks.com/matlabcentral/fileexchange/48238-nonstationary-extreme value-analysis-neva-toolbox}{\soft{Matlab} package} \pkg{NEVA} uses a differential evolution Markov chain algorithm for estimating univariate nonstationary models \citep{Cheng:2014}.

\begin{table}
\centering
\begin{tabular}{lllllll}
\toprule
package & function & models & covariates & sampling & prior choice\\
\midrule
\pkg{evdbayes} & \code{posterior} & 1--4 & loc./thresh & RWMH & multiple  \\
\pkg{extRemes} & \code{fevd} & 1--4,* & all & RWMH & custom  \\
\pkg{INLA} & \code{inla} & 1--2,* & loc./thresh & -- & PC  \\
\pkg{MCMC4Extremes} & \code{ggev}, \ldots
& 1--2,* & no & RWMH & fixed \\
\pkg{revdbayes} & \code{rpost} & 1--4 & no & RU & custom
\\
 \pkg{texmex}& \code{evm} & 1--2,* & all & IMH & Gaussian  
\\
\bottomrule
\end{tabular}

\caption{Comparison of \soft{R} packages for Bayesian univariate extreme value modeling. Families: generalized extreme value distribution (1), generalized Pareto distribution (2), inhomogeneous Poisson process (3), order statistics/$r$-largest (4) or custom/other (*). Sampling: random walk Metropolis--Hastings (RWMH), exact sampling ratio-of-uniforms (RU), independent Metropolis--Hastings (IMH); the \pkg{INLA} package uses deterministic Laplace approximations. ``PC''  priors refer to penalized complexity priors. All packages, except \pkg{evdbayes}, also provide \code{S3} methods (notably \code{plot} and \code{summary}).  All packages return a matrix of posterior draws.}
\label{tab:bayesian}
\end{table}

\subsection{Semiparametric inference for univariate extremes} \label{sec:nonpar}

In the semiparametric approach to extremes,

some components of the probability structure are handled through a relatively general (and nonparametric) asymptotic structure, which can be extrapolated towards higher yet unobserved quantile levels, for instance for the purpose of extreme-quantile estimation. The parametric form includes the shape parameter $\xi$ and potentially second-order regular variation indices, $\rho$. \cite{Caeiro2016} provides a review of many estimators discussed next with an emphasis on the choice of the number of order statistics to keep for inference, which has close ties to threshold selection methods discussed in \Cref{sec:threshselect}.

Consider a sample of independent and identically distributed variables $X_1, \ldots, X_n$ with quantile function $Q(x)$.
Assuming heavy-tailed distributions with limiting shape parameter $\xi>0$, the survival function $S(x) = x^{-1/\xi} L_F(x)$ if and only if $Q(1-1/x) = x^\xi L_U(x)$,
with $\xi$ the shape parameter, and $L_F$ and $L_U$  slowly varying functions, meaning $\lim_{x\to \infty} L(tx)/L(x)=1$ for any $t>0$ \citep[][\S~5]{Ledford:1996}.

Nonparametric estimators of the extreme value index are widespread, most of them variants of the \cite{hill75} estimator. Let $X_{1,n} \leq \cdots \leq X_{n,n}$ be the order statistics of the sample of size $n$: the Hill estimator is the mean
excess value of log-transformed data of the $k$ largest values,
\begin{align}
H_{k,n} = \frac{1}{k} \sum_{j=1}^k \log\left(\frac{X_{n-j+1,n}}{X_{n-k,n}}\right).
\label{eq:Hill:estimator}
\end{align}
The Hill estimator is generally computed for a wide range of values of $k$, which leads to so-called Hill plots $(k, H_{k,n})$, $k=1, \ldots, n$.
Under certain regularity conditions, the Hill estimator \eqref{eq:Hill:estimator} is asymptotically normal \citep[e.g.,][Section~4.4]{beirlantetal04}.
A large number of \soft{R} packages provides functions to estimate  \eqref{eq:Hill:estimator} and to make
Hill plots such as \pkg{evir} \citep{Pfaff:2018aa},
\pkg{evmix} \citep{Hu:2018aa},
\pkg{extremefit} \citep{Durrieu:2018aa},
\pkg{ExtremeRisks} \citep{Padoan:2020aa},
\pkg{ptsuite} \citep{Munasinghe:2019aa},
\pkg{QRM} \citep{Pfaff:2020aa},
\pkg{ReIns} \citep{Reynkens:2020aa} and \pkg{tea} \citep{tea:2020}.

The performance of the Hill estimator strongly depends on the number of observations kept to estimate the tail index: $H_{k,n}$ has a large variance if $k$ is too small, whereas the Pareto-type tail behavior might not be verified for the selected $k$ largest values if $k$ is too large. The choice of $k$ is typically based either on an empirical rule to find the area where $H_{k,n}$ is ``stable''
or by minimizing the asymptotic mean squared error (AMSE).
A large number of those algorithms to minimize the latter are provided in \pkg{tea} along with the bootstrap methods of \cite{Hall/Welsh:1985}, \cite{Hall:1990}, \cite{Danielsson:2001}, \cite{Caeiro:2014} and \cite{Caeiro2016}.

The Hill estimator \eqref{eq:Hill:estimator} has a number of drawbacks: it supposes $\xi>0$, it is not consistent, is not translation invariant and it behaves erratically for large $k$.
Many alternative estimators try to palliate these lacks.

Since the Hill estimator has nondifferentiable sample paths with respect to the threshold value, the choice of threshold is notoriously difficult. \cite{resnick1997smoothing} proposed a smoothed version of the Hill estimator based on averaging consecutive estimates via a moving window; these plots are provided in \pkg{evmix} and \pkg{tea}. The random block maximum estimator \citep{Wager:2014}, constructed as a $U$ statistic, has infinitely differentiable sample paths and is thus much less sensitive to the choice of $k$ than most Hill-type estimators. The unpublished package \href{https://github.com/swager/rbm}{\pkg{rbm}} is available from Github.

Packages \pkg{evt0} \citep{Manjunath:2013aa}
and \pkg{ReIns} implement the generalized Hill estimator based on a uniform kernel estimation
\citep{beirlant1996excess}.

\pkg{evt0} also provides functions for the location-scale invariant version of the Hill estimator
introduced by \cite{santosalvesgomes:2006} and the biased-reduced version of  \cite{Fernanda2012}, as well as a mixed moment estimator and location invariant alternative.

The package \pkg{extremefit} implements the kernel-weighted version of the Hill estimator of
\cite{grama2008statistics};

the authors provide an automatic selection procedure for the threshold $u$, with functions to handle these weighted estimators either for user-supplied weights or for weights automatically selected using an adaptative selection.

\subsubsection{Moment estimators and other alternative estimators}

While maximum likelihood estimation and Hill-type estimators are most commonly used for the shape parameter, other estimators are available and may be more robust in small samples. One such was proposed by \cite{dekkers1989moment} and  \pkg{evt0} provides a generalization of the latter by \cite{santosalvesgomes:2006}. Since moments of extreme value distributions may not exist if $\xi>0$, we can consider instead a bijection between the parameter vector $\mathbf{\theta}$ and probability weighted moments of the form $\mathsf{E}[X^p F(X)^q\{1-F(X)\}^r]$ for integers $p, q, r$ \citep{Hosking/Wallis:1987}. Another avenue is to match sample linear combinations of order statistics with their theoretical counterparts using (trimmed) $L$-moments \citep{Hosking:1990}. A group of \soft{R} packages, including \pkg{lmom} \citep{lmom}, \pkg{lmomco} \citep{lmomco}, \pkg{TLMoments} \citep{TLMoments} implement these approaches for a variety of common distributions (as does the \soft{Python} package \pkg{lmoments}), but some also allow custom distribution functions. \pkg{extRemes} also implements $L$-moments, while \pkg{RobExtremes} \citep{RobExtremes} provides robust estimators of the extreme value parameters and \pkg{laeken} \citep{Alfons.Templ:2013} proposes robust modelling of Pareto data. Package \pkg{extremeStat} \citep{extremeStat} includes functionalities to compute extreme quantiles based on $L$-moments estimator.

\subsubsection{Quantile, expectile and extremiles}
In the heavy-tailed setting with $\xi>0$, \cite{weissman78} proposed estimating the tail quantile $Q(1-p)$ for given small tail probability $p$ using
\begin{align*}
q^W_{k,n}(p) = X_{n-k,n} \left\{ \frac{k+1}{p(n+1)} \right\}^{H_{k,n}},                                                                     \end{align*}
where $H_{k,n}$ is the Hill estimator \eqref{eq:Hill:estimator}. \pkg{ReIns} implements the Weissman estimator either specified by the probability level $p$ or by the return period $1/p$. The Weissman-type estimator for the class of estimators proposed by \cite{santosalvesgomes:2006}
are provided by \pkg{evt0}, whereas \soft{R} package \pkg{extremefit} gives the quantile corresponding
to weighted Hill estimator. Bias-corrected versions of the Weissman estimator also exist, yet are seemingly not implemented in software.

Quantiles can be formulated as the solution of an asymmetric piecewise linear loss function. Taking instead an asymmetric quadratic loss function yields expectiles \citep{Newey/Powell:1987}, another risk measure gaining popularity in risk management \citep{Bellini.diBernardino:2017}. Many recent work studies their extremal property: on the software side, \pkg{ExtremeRisks} implements the methodology of \cite{Davison.Padoan.Stupfler:2021,Padoan.Stupfler:2022}, including estimation of expectiles using Hill-type estimators, test of equality of tail expectiles and confidence regions for extreme expectiles. An alternative risk measures, the so-called extremiles, has been developed recently \citep[e.g., ][]{Daouia.Gijbels.Stupfler:2022} but no software implementation exists at the time of writing.

\begin{table}[htb!]
\begin{tabular}{llll}
\toprule
package & estimation &  function & features \\
\midrule
\pkg{evir} & --- &  \code{hill} & \textbf{e},\textbf{p} \\
\pkg{evmix} & smoothing & \code{hillplot} & \textbf{p} \\
\pkg{evt0} & location invariant &  \code{gh}, \code{PORT.Hill} & \textbf{p}, \textbf{q} \\
\pkg{extremefit} & weighted, time series & \code{hill}, \code{hill.adapt}, \code{hill.ts} & \textbf{e}, \textbf{p}, \textbf{q}, \textbf{o}\\ 
\pkg{ExtremeRisks} & time series, CI&  \code{HTailIndex}, \code{EBTailIndex} &  \textbf{e}, \textbf{o}\\
\pkg{fExtremes} & --- & \code{hillPlot}, \code{shaparmHill} & \textbf{e}, \textbf{p} \\
\pkg{ptsuite} & --- & \code{alpha\_hills} & \textbf{e}\\
\pkg{QRM} & --- & \code{hill}, \code{hillPlot} & \textbf{e}, \textbf{p}\\
\pkg{rbm} & random block & \code{rbm}, \code{rbm.plot} & \textbf{e}, \textbf{p} \\
\pkg{ReIns} &conditional, censoring & (\code{c})\code{Hill}, (\code{c})\code{genHill}, \code{crHill}, \ldots & \textbf{e}, \textbf{p}\\
\pkg{tea} & smoothing & \code{althill}, \code{avhill} & \textbf{p} \\
\bottomrule
\end{tabular}
\caption{Main functionalities of \soft{R} packages for nonparametric Hill-type estimators of the shape parameter, including functionalities for estimation of the shape or tail index (\textbf{e}), Hill threshold diagnostic plots (\textbf{p}), quantile estimates (\textbf{q}) and other methods (\textbf{o}).}
\label{tab:hillplots}
\end{table}

\subsection{Threshold selection} \label{sec:threshselect}

Many methods are driven by analyses of the most extreme observations.  In the univariate case, these are the $k$ largest order statistics or, equivalently, observations that exceed a threshold $u$ as presented in the previous section.  The underlying theory considers limiting behavior as the threshold increases.  In practice, a suitably high threshold is set empirically, balancing the bias from using a low threshold that violates the theory with statistical imprecision from using a threshold that is unnecessarily high.   For information about many of the following methods, see the review of \cite{Scarrott2012}. Methods for semiparametric estimators based on variants of Hill's estimator for the shape were presented in \Cref{sec:nonpar}.

\subsubsection{Visual threshold selection diagnostics}

In a \emph{threshold stability plot}, point and interval estimates of parameters are plotted against a range of threshold values.  A particular example is the \emph{Hill plot} featured in \Cref{sec:nonpar} (see Table \ref{tab:hillplots} for an overview of available implementations).  In the univariate case, the focus is often on the shape parameter $\xi$: we choose the lowest threshold above which we judge the point estimates of $\xi$ to be approximately constant in threshold, bearing in mind statistical uncertainty quantified by the interval estimates.  These inferences may be based on the generalized Pareto distribution \eqref{eq:gpdist} for threshold excesses or the inhomogeneous Poisson process model \eqref{eq:pp_conv}, using a frequentist or Bayesian analysis.  In the generalized Pareto case, the threshold-independent scale parameter $\sigma_u^* = \sigma_u - \xi u$ is used. In the frequentist case, it is useful to have the option to calculate the intervals using profile likelihoods, because they tend to have better coverage properties than Wald intervals, especially for high thresholds.

If a generalized Pareto distribution with $\xi <1$ applies at threshold $u$ then the mean excess ${\rm E}(X - v \mid X > v)$ is a linear function of $v$ for all $v > u$.  This motivates the \emph{mean residual life (MRL) plot}, in which the sample mean of excesses of a range of thresholds are plotted against the threshold, with pointwise confidence intervals superimposed.  We choose the lowest threshold above which the plot appears linear. Table \ref{tab:thresholdgraphs} summarises the functionality of \soft{R} packages in terms of these plots.

\begin{table}
\centering
\begin{tabular}{lllclll}
\toprule
package & stability & models & profile & inference & MRL \\
\midrule
\pkg{eva} & \code{gpdDiag} & 1\hphantom{,2} & yes & \textsc{mle} & \code{mrlPlot} \\
\pkg{evd} & \code{tcplot} & 1,2 & no & \textsc{mle}/\textsc{b} & \code{mrlplot} \\
\pkg{evir} & \code{shape} & 1\hphantom{,2} & no & \textsc{mle} & \code{meplot} \\
\pkg{evmix} &  \code{tcplot} & 1\hphantom{,2} & no & \textsc{mle} & \code{mrlplot} \\
\pkg{extRemes} & \code{threshrange.plot} & 1,2 & no & \textsc{mle} & \code{mrlplot}\\
\pkg{fExtremes} & \code{gpdShapePlot}, \ldots & 1\hphantom{,2} & no & \textsc{mle} & \code{mrlPlot}\\
\pkg{ismev} & \code{gpd.fitrange}, \code{pp.fitrange} & 1,2 & no & \textsc{mle} & \code{mrl.plot} \\
\pkg{mev} & \code{tstab.egp}, \code{tstab.gpd} & 1,3 & yes & \textsc{mle}/\textsc{b} & \code{automrl} \\
\pkg{POT} & \code{tcplot} & 1\hphantom{,2} & no & \textsc{mle} & \code{mrlplot} \\
\pkg{QRM} &  \code{xiplot} &  1\hphantom{,2} & no & \textsc{mle} & \code{MEplot} \\
\pkg{ReIns} & \code{1Dmle} & 1\hphantom{,2} & ---  &  \textsc{mle} & \code{MeanExcess} \\
\pkg{texmex} & \code{egp3RangeFit}, \code{gpdRangeFit}& 1,3 & no & \textsc{mle}/\textsc{b} & \code{mrl}\\
\pkg{threshr} & \code{stability} & 1\hphantom{,2} & yes & \textsc{mle} & --- \\
\bottomrule
\end{tabular}
\caption{Comparison of \soft{R} packages for classical visual methods.  Stability: function name for a threshold stability plot; models: either generalized Pareto (1),  inhomogeneous Poisson process (2) or extended generalized Pareto model of \cite{Papa2013} (3); inference: method of inference, either maximum likelihood estimation (\textsc{mle}) or Bayesian (\textsc{b}); MRL: mean residual life plot, if applicable. }
\label{tab:thresholdgraphs}
\end{table}

The \code{lmomplot} function in the \pkg{POT} package can help to identify for which thresholds the sample $L$-skewness and $L$-kurtosis of excesses are related as expected under a generalized Pareto distribution.  These plots require the use of subjective judgement to select a threshold.  More formal methods seek to reduce subjectivity and perhaps introduce a greater degree of automation.

\subsubsection{More formal methods}

\noindent \emph{Penultimate models}. Formal testing procedures compare the null hypothesis of having a generalized Pareto distribution above a threshold $u$ against an alternative model. Theoretically-justified alternative models can be derived from the penultimate approximation to extremes, either by selecting piecewise constant shape \citep{Northrop2014} or by using tilting function to provide more general models that should have faster convergence. The models proposed in \cite{Papa2013} lead to a threshold stability plot for an additional parameters. These approaches are implemented in \pkg{mev}.
\vspace{0.25cm}

\noindent \emph{Goodness-of-fit diagnostics}. One drawback of the threshold stability plot and tests is that they do not entirely indicate whether the tail model fits the data well. Goodness-of-fit diagnostics can thus complement other diagnostics. The \pkg{eva} package \citep{eva} provides multiple testing methods with the Cram\'{e}r--von Mises and Anderson--Darling criteria and Moran's tests, all with control for the false discovery rate \citep{Bader2018}. The benefit of this approach, compared to visual diagnostics, is that it does not require user input and is more readily implementable with large multivariate or spatial data sets. The approach of \cite{Dupuis:1999}, based on examination of the weights attached to the largest observations from the sample and obtained using a robust fitting procedure, can be obtained via \pkg{mev}.
\vspace{0.25cm}

\noindent \emph{Sequential analysis and changepoints}. Parameter estimates obtained by fitting a tail model at multiple consecutive thresholds are dependent because of the non-negligible sample overlap. The \pkg{mev} package provides the method of \cite{Wadsworth2016}, which exploits a technique from sequential analysis by fitting a point process over a range of thresholds and building an approximate white noise sequence from the differences between consecutive estimates using their asymptotic covariance matrix, suitably rescaled to be standard normal. The \pkg{tea} package provides the  Pearson $\chi^2$ test of normality applied to sequences of differences of scale estimates, following \cite{Thompson:2009}, while threshold stability plots based on estimates of the coefficient of variation and sequential testing of \cite{Castillo/Padilla:2016} are included in \pkg{ercv}.

Gerstengarbe--Werner plots \citep[cf.,][Appendix~B]{Cebrian2003} are graphical diagnostic plots derived from consecutive differences between order statistics. Sequential Mann--Kendall tests are performed from smallest to largest order statistics (and vice-versa), with the intersection point used as a changepoint candidate; such plots can be created with \pkg{tea}.
\vspace{0.25cm}

\noindent \emph{Predictive performance}. The \pkg{threshr} package \citep{Northrop/Attalides/Jonathan:2017} looks at the predictive performance of the generalized Pareto for a binomial-generalized Pareto model fitted using the Bayesian approach. The scheme uses a leave-one-out cross validation scheme for values at a fixed validation threshold $v$ at or above the range of potential thresholds considered.
\vspace{0.25cm}

\noindent \emph{Mixture models}. The generalized Pareto specifies a distribution only for exceedances above a threshold $u$, but having a model below this threshold may be desirable, with some options enabling automatic threshold selection. The \pkg{evmix} package \citep{Hu:2018aa} provides implementations of most of the mixture models listed in \cite{Scarrott2012}: this includes parametric models for the bulk of the data (for which users can inform threshold selection by looking at the profile likelihood for $u$),  nonparametric and kernel-based approaches for the data below the threshold. Many such models are discontinuous at the threshold and require choosing a fixed threshold. The \pkg{extremefit} package \citep{Durrieu:2018aa} provides a mixture model implementation with a kernel-based bulk model and adaptive selection rules for the bandwidth parameter. The \pkg{mev} package provides the extended generalized Pareto model of \cite{Naveau:2016} for modeling rainfall.
\vspace{0.25cm}

\begin{table}
\centering
\begin{tabular}{llll}
\toprule
type & methods & package & function  \\
\midrule
 penultimate & \cite{Northrop2014} & \pkg{mev} & \code{NC.diag} \\
  & \cite{Papa2013} & \pkg{mev} & \code{tstab.egp}\\
goodness-of-fit & \cite{Gerstengarbe:1989} & \pkg{tea} & \code{ggplot} \\
 & \cite{Hosking/Wallis:1997} &\pkg{POT} & \code{lmomplot} \\
 & \cite{Bader2018} & \pkg{eva} & \code{gpdSeqTests} \\
sequential &  \cite{Wadsworth2016} & \pkg{mev} &  \code{W.diag}\\
& \cite{Thompson:2009} & \pkg{tea} &  \code{TH} \\
& \cite{Castillo/Padilla:2016} & \pkg{ercv} & \code{cvplot}, \code{thrselect}\\
predictive & \cite{Northrop/Attalides/Jonathan:2017} & \pkg{threshr} & \code{ithresh} \\
mixture & \cite{Scarrott2012} & \pkg{evmix} & ---  \\
 & \cite{Durrieu2015} & \pkg{extremefit} &$\cdot$\code{paretomix} \\
 & \cite{Naveau:2016} & \pkg{mev} & \code{fit.extgp}\\
\bottomrule
\end{tabular}
\caption{Overview of formal threshold selection methods and numerical implementations}
\label{tab:formalthreshold}
\end{table}

 \subsection*{\textbf{Univariate extremes implementations in other programming languages}}
While \soft{R} is arguably the programming language boasting the most software implementations used for extreme value analyses, some basic routines are available elsewhere for estimation of univariate models using maximum likelihood or probability weighted moments: these include the \soft{Julia} package \href{https://juliapackages.com/p/extremes}{\pkg{Extremes}}, the \soft{Matlab} \pkg{EVIM} package and the \soft{Python} packages \code{wafo}, \pkg{pyextremes} and \pkg{scikit-extremes}.

\section{Multivariate extremes}\label{sec:multivariate}
The lack of ordering of $\mathbb{R}^D$ leads to multiple definitions of extremes \citep{Barnett:1976}. We focus on componentwise maxima and concomitant exceedances, which lead to the multivariate analog of block maximum and peaks over threshold methods. Another option, structure variables, reduces the data to univariate summaries and can be dealt with using tools presented before.

\subsection{Multivariate maxima}
 Consider independent and identically distributed sequence of $D$-variate random vectors $\{\boldsymbol{Y}_i\}_{i \geq 1}$, where each vector $\boldsymbol{Y}_i$ has marginal distribution functions $F_j$ $(j=1, \ldots, D)$. Paralleling the univariate case, we consider the vector of componentwise maxima $\boldsymbol{M}_n=(M_{n,1}, \ldots, M_{n,D})$, where $M_{n,j} = \max\{Y_{1,j}, \ldots, Y_{n, j}\}$. If there exists sequences of location and scale vectors $\boldsymbol{a}_n\in \mathbb{R}^D_{+}$ and $\boldsymbol{b}_n \in \mathbb{R}^D$ such that
\begin{align*}
\lim_{n \to \infty}\Pr\left\{\boldsymbol{a}_n^{-1}(\boldsymbol{M}_n-\boldsymbol{b}_n)\right\}
=G(\boldsymbol{y}),
\end{align*}
for $G$ non-degenerate, then the limiting distribution is of the form
\begin{align}
 G(\boldsymbol{y}) = \exp \left\{ - D \int_{\mathbb{S}_D} \max\left(\frac{\boldsymbol{w}}{\boldsymbol{z}}\right) \mathrm{d}H(\boldsymbol{w})\right\},
 \label{eq:bivariate:Glimit:1}
\end{align}

where the so-called spectral measure $H$ is a probability measure on the $D$-simplex $\mathbb{S}_D=\{\boldsymbol{\omega} \in \mathbb{R}_{+}^D: \|\boldsymbol{\omega}\|_1=1\}$ and the margins are generalized extreme value distributed.
Distributions of the form \cref{eq:bivariate:Glimit:1} are termed multivariate extreme value, or max-stable. The simple max-stable distribution is obtained upon setting $\mu=1, \sigma=1, \xi=1$ for each margin, corresponding to the unit Fréchet.

The lack of specification for $H$, other than the $D$ moment constraints $\mathsf{E}(S_j) = 1/D$ $(j=1,\ldots, D)$ for $\boldsymbol{S} \sim H$, means that the set of probability measures satisfying the moment constraint is infinite, unlike in the univariate case. The \pkg{copula} package includes three tests of the max-stability assumption; see \cite{Kojadinovic/Segers/Yan:2011,Kojadinovic/Yan:2010,Ghorbal/Genest/Neslehova:2009}, while the graphical diagnostic proposed by \cite{Gabda:2012} is part of \pkg{mev}.

\noindent \emph{Likelihood-based estimation}. The likelihood of a simple max-stable random vector $\boldsymbol{Z}$ with a parametric model for $V$ is obtained by differentiating the distribution function $\exp\{-V(\boldsymbol{z})\}$ with respect to each $z_1, \ldots, z_D$.

The number of terms in the likelihood is the $D$th Bell number, which is the total number of partitions of $D$ elements into $k$ $(k=1, \ldots, D)$ elements. Even in moderate dimensions, the number of distinct likelihood contributions is huge and the calculations become prohibitive. One way to circumvent this problem is to add the information about the partition if occurrence times are recorded \citep{Stephenson:2005}; this is implemented in the \pkg{evd}, but only for bivariate models. In practice, we replace the limiting partition with the empirical one. The likelihood is biased unless $n \gg D$ since the empirical partition also needs to converge to the limiting hitting scenario; for weakly dependent processes, use of the observed partition may induce bias
\citep{Wadsworth:2015}. Instead, \cite{Thibaud:2016} propose to impute the partition using a Gibbs sampler, while \cite{Huser/Dombry/Ribatet/Genton:2019} use a stochastic expectation-maximisation algorithm; the $E$-step for the missing partition uses a Monte-Carlo estimator, where approximate draws are obtained from the Gibbs sampler of \cite{Dombry/Eyi-Minko/Ribatet:2013}. None of these extensions have been implemented in publicly available software packages.

\noindent \emph{Parametric models}. While max-stable models have been around for a while, there are few software implementations for estimating such models. The \pkg{evd} and \pkg{copula} packages provide functionalities that are restricted to the bivariate setting. The \pkg{SpatialExtremes} and \pkg{CompRandFld} packages have methods for fitting max-stable processes using pairwise composite likelihood for spatial models; see Section~\ref{sec:functional}.

There are only handful of useful parametric models that generalize to dimension $D>2$. The prime example is the logistic multivariate extreme value model, which is overly simplistic and lacks flexibility since the distribution is exchangeable. Many existing models are special cases of a Dirichlet family of distributions \citep{Belzile:2017} and obtained through tilting \citep{Coles:1991} to satisfy the moment constraint. These all have the drawback that the number of parameters is constant or grows linearly with the number of dimensions $\mathrm{O}(D)$ and this typically isn't enough  for characterizing complex data. Two models derived from elliptical distributions, the  H\"usler--Reiss model \citep{Huesler:1989} and the extremal Student-$t$ \citep{Nikoloulopoulos:2009}, are more useful in large dimensions because their scale matrix can be used to parametrize the pairwise dependence individually with $\mathrm{O}(D^2)$ entries, and they can be more readily adapted to the functional setting, with extensions for skew-symmetric families \citep{Beranger:2017}. The last parametric family, of which the most prominent example is the asymmetric logistic distribution, are max-mixtures \citep{Stephenson:2005} that assign different weights to multiple simultaneous combinations of extremes. This allows for some degree of asymmetry and asymptotic independence, but such models are  overparametrized with $\mathrm{O}(2^D)$ coefficients.

Joint estimation of all marginal and dependence parameters is complicated because of the potential high-dimensionality of the optimization problem, but also because of potential model misspecification that leads to unplausible parameter estimates. It is therefore common to use a two-stage approach, whereby data are first transformed to standardized margins

and then  dependence parameters are fitted separately. The function \code{fbvevd} in \pkg{evd} allows the user to pass fixed values for some parameters. The \pkg{tailDepFun} package contains routines for fitting the continuous updating weighted least squares estimator, along with goodness-of-fit tests, for multivariate and functional models including max-linear models \citep{Einmahl/Segers/Kiriliouk:2018}.

A different avenue is to estimate an equivalent form of $H$ termed the Pickands dependence function \citep[cf.,][p. 150]{Falk:2011}, which is equivalent to \cref{eq:bivariate:Glimit:1}.

Enforcing properties of the Pickands dependence function, notably convexity and known values on the corners of the simplex, can improve estimation.

The \pkg{evd} package  allows users to estimate nonparametrically the bivariate dependence function based on the estimators of \cite{pickands81} and \cite{caperaafougeregenest97} for block maxima; additional options correct for boundary and convexity constraints.

While most estimators are restricted to the bivariate case, \pkg{copula} with its function \code{An} provides generalization to higher dimensions for multiple estimators \citep{Gudendorf/Segers:2012}. Multivariate estimators based on Bernstein polynomials that guarantee convexity \citep{Marcon:2017} are provided by the \code{beet} procedure in \pkg{ExtremalDep}, along with the \code{madogram} estimator. Bayesian estimation is also available in the bivariate case, imposing a prior on the order of the Bernstein polynomials. The package also includes a procedure for computing pointwise confidence intervals using a nonparametric bootstrap.
The \code{UniExtQ} from \pkg{ExtremalDep} provides credible intervals for bivariate extreme quantile regions \citep{Beranger:2021}, estimated using an extension of this approach. Lastly, \pkg{fCopulae} provides parametric dependence function, correlation coefficient and tail dependence measures for bivariate extreme value copulas.

\noindent \emph{Unconditional simulation algorithms}.
For a long time, exact unconditional simulation algorithms for max-stable processes were elusive outside of special cases \citep{Schlather:2002}. Both \pkg{mev} and  \pkg{graphicalExtremes} implement the algorithm of \cite{Dombry:2016} for selected multivariate models (including for the latter extremal graphical models on trees) ensuring exact simulation, whereas  \pkg{evd} uses dedicated algorithms for logistic and asymmetric logistic models in arbitrary dimensions \citep{Stephenson:2003}.  The  \pkg{copula} (\code{evCopula} objects) \citep{Yan:2007} and \pkg{SimCop} packages \citep{Tajvidi/Turlach:2018} have functionalities for simulation of some bivariate extreme value distributions and the multivariate logistic model, or Gumbel copula. Packages \pkg{mev} and \pkg{BMAmevt} provide random number generators for selected parametric angular density models.

\subsection{Threshold models} \label{sec:mvthresh} 

Multivariate regular variation, which underlies the max-stable distribution of \Cref{eq:bivariate:Glimit:1}, can also be used for threshold exceedances by considering the associated Poisson point process of extremes with intensity measure $\Lambda$ on a risk region $\mathcal{R} \subset \mathbb{R}^D_{+} \setminus \{\boldsymbol{0}_D\}$, i.e., the positive orthant excluding the origin \citep{Resnick:1987}. Assuming the intensity measure is absolutely continuous, the intensity function $\lambda(\boldsymbol{x}) = \partial^D \Lambda(\boldsymbol{x})/(\partial x_1 \cdots \partial x_D)$  exists and we can define a density over $\mathcal{R}$ by renormalizing $\lambda(\boldsymbol{x})$ by the measure of the risk region $\Lambda(\mathcal{R})$. The resulting likelihoods of the point process, multivariate generalized Pareto distributions and more general threshold models are much simpler than their max-stable counterpart, but there are typically two numerical bottlenecks associated to fitting these models. The first arises from the calculation of the measure of the risk region, which is often intractable and  must thus be estimated using Monte Carlo methods. There are closed-form expressions for the special case $\mathcal{R}=\{\boldsymbol{x} \in \mathbb{R}^{D}_{+}: x_i >u\}$; if $\xi = \mathbf{1}_D$, then $\mathcal{R}=\{\boldsymbol{x} \in \mathbb{R}^{D}_{+}: \|\boldsymbol{X}\|_1 >u\}$ has risk measure $\Lambda(\mathcal{R})=Du^{-1}$ irrespective of the model for $\Lambda$. The second bottleneck is due to censoring: not all components of a random vector may be extreme and the limiting model may be a poor approximation at finite levels for weakly dependent vectors \citep{Ledford:1996}. To reduce the bias arising from consideration of the asymptotic distribution, it is customary to left-censor observations falling below marginal thresholds. Most multivariate peaks over threshold models  are based on the multivariate generalized Pareto \citep{Tajvidi:2006}, whereby $\mathcal{R} = \{\boldsymbol{x} \in \mathbb{R}^D_{+}: \max_{j=1}^D x_j > u\}$ --- other constructions are described in \cite{Rootzen:2018}. \cite{Kiriliouk:2019} provide expressions for the likelihood of many parametric models with strategies for diagnostics; these are not currently implemented in software. The point process likelihood can  also be used in place of the multivariate generalized Pareto: in the bivariate case, the \pkg{evd} package proposes the point process likelihood \citep{Smith:1997}, but the bivariate censored likelihood implemented therein actually uses the max-stable copula instead \citep{Ledford:1996}.

Most implementations are restricted to the bivariate setting or are reserved for spatial data. The \pkg{graphicalExtremes} package \citep{Engelke/Hitz:2020} is a notable exception: it implements the multivariate H\"usler--Reiss generalized Pareto distribution for graphical models. Exploiting the relation between the model and conditional extremal dependence, the parameters of the H\"usler--Reiss or Brown--Resnick process are directly related to the variogram matrix, whose entries are estimated empirically using pairwise empirical estimators of $\chi$. The full likelihood can be used (including censoring), but the factorization of the likelihood over cliques allows for higher-dimensional models to be fitted through maximum likelihood at reasonable cost, since each component is low dimensional. The \code{bvtcplot} function in the \pkg{evd} package provides threshold stability plots in the bivariate case based on the spectral measure.  \pkg{mev} provides composition sampling algorithms for threshold models for various risk functionals $\mathcal{R}$ in the multivariate setting \citep{Ho:2019}.

Rather than condition on the maximum component exceeding a threshold, we can focus instead on exceedances of the $j$th component, i.e., consider a limiting model for $\boldsymbol{Y}_{-j} \mid Y_j > u$.  \cite{Heffernan:2004} showed that a particular choice of normalizing sequences allows for the existence of non-degenerate limiting measure, including for asymptotically independent models. Inference for the  conditional extremes model is usually performed  in two stages. In the first, the marginal distributions are estimated semiparametrically and data are transformed to Laplace margins \citep{Keef:2013}. In the second step, the dependence parameter vectors are estimated using a nonlinear regression model under the assumption of Gaussian residuals.

 Inference for the conditional extremes model as implemented in the \pkg{texmex} package relies on simulation: 
 the probability of extreme events is obtained by calculating the fraction of simulated points falling in the risk region and uncertainty quantification is done using the bootstrap scheme described in \cite{Heffernan:2004}.

The multivariate regular variation representation provides another modeling approach for peaks over threshold using radial exceedances. For this, the data are first transformed to standardized scale with unit shape and the resulting components $\boldsymbol{Y}$ are mapped to pseudo-angles $(R, \boldsymbol{\Theta})$, with, e.g., $R=\|\boldsymbol{Y}\|_1$ and $\boldsymbol{\Theta} = \boldsymbol{Y}_{-D}/R$.  Since $R$ and $\boldsymbol{\Theta}$ become stochastically independent  as $R$ tends to infinity, one can focus on modelling the spectral measure $H(\boldsymbol{\theta})$ appearing in \cref{eq:bivariate:Glimit:1}.    
 \pkg{ExtremalDep} supports composite likelihood maximum estimation with pseudo-angles for $D$-dimensional distributions, with composite likelihood information criterion estimates to compare models, with density functions, plots and associated functions for each parametric model. Nonparametric estimation of the spectral measure only requires the user to impose mean contraints. Starting from a sample of pseudo-angles, these can be enforced through empirical likelihood method \citep{Einmahl/Segers:2009} or Euclidean likelihood \citep{deCarvalho:2013}. The \pkg{extremis} package \citep{extremis} implements these functionalities in the bivariate setting, and \pkg{mev} 
 in higher dimensions. The unpublished \href{https://www.lancaster.ac.uk/~wadswojl/}{\pkg{EVcopula} package} implements the bivariate model of \cite{Wadsworth2016} along with likelihood-based estimation methods and can be used to estimate probabilities of large bivariate quantiles for both asymptotic (in)dependence scenarios. The \pkg{BMAmevt} package is dedicated to the implementation of a Bayesian nonparametric model that uses a trans-dimensional Metropolis algorithm for fitting a Dirichlet mixture to the spectral measure based on pseudo-angles in moderate dimensions \citep{Sabourin:2014}.

\subsection{Coefficients of tail dependence and structural variables}

In multivariate settings, knowing the speed of decay of the dependence between pairs of random variable is useful for risk assessment. This also helps validate empirically if asymptotic multivariate extreme value models that assume multivariate regular variation are warranted or not. The tail correlation coefficient  \citep{Coles:1999},
\begin{align}
\chi(v) =  \frac{\Pr[\min_i \{F_i(Y_i) >v\}]}{1-v}, \label{chiD}
\end{align}
with $\chi=\lim_{v \to 1} \chi(v)$, is used to assess whether extremes are asymptotically independent ($\chi=0$) or dependent ($\chi>0$). \Cref{chiD} suggests replacing the unknown distribution functions by their empirical counterpart. The bivariate empirical estimator is usually defined as $2-\log\{\widehat{C}(v\boldsymbol{1}_2)/\log(v)$ rather than $\{1-C(v,v)\}/(1-v)$ by making the approximation $1-v \sim -\log(v), 1-C(v,v)\sim -\log\{C(v,v)\}$ for $v \ll 1$.

A related coefficient measuring dependence is the coefficient of tail dependence, often denoted $\eta$. With data transformed to unit Pareto margins, say $\boldsymbol{Y}^{\mathrm{p}}$, the structural variable $T^{\mathrm{p}} =\min_{j=1}^D Y_j^{\mathrm{p}}$ is such that, for large $u $ \citep[][eq.~5.6]{Ledford:1996},
\begin{align}
	\Pr\left(T^{\mathrm{p}}>u+t \mid T^{\mathrm{p}}>u\right) \sim \frac{L(u+t)}{L(u)} (1+t/u)^{-1/\eta}, \label{eq:coef-tail-dep}
\end{align}
with $L(x)$ a slowly varying function.
 The $\eta$ coefficient can be estimated  by fitting a generalized Pareto distribution with shape $\eta$ and scale $\eta u$ to exceedances of $T^{\mathrm{p}}$ above $u$. If data are transformed to the exponential scale instead, the scale parameter of the structural variable is $\eta$ and the maximum likelihood estimator of the latter coincides with Hill's estimator (\Cref{sec:nonpar}).

The coefficient of tail dependence takes values in $(0,D^{-1})$ if the variables are negatively associated, $\eta=D^{-1}$ for independent variables, and $\eta \in(D^{-1},1]$ if the variables exhibit positive association.
 In the multivariate setting, the coefficients $\eta_C$ for  subsets $C \subset \{1, \ldots, D\}$ satisfy ordering constraints with $\eta_{C} \leq \eta$ \citep{Schlather:2003}.

In the bivariate setting, it is customary to consider $\bar{\chi}=2\eta-1$ instead of $\eta$, which gives $\bar{\chi} \in (-1,1]$ \citep{Coles:1999}. The \pkg{evd} package function \code{chiplot} provides plots of $\chi$ and $\bar{\chi}$ based on the empirical distribution of the minimum, with approximate pointwise standard errors through the delta-method. The \pkg{mev}  package provides  various estimators of $\eta$ and $\chi$, while \pkg{graphicalExtremes} includes empirical estimators \code{emp\_chi} that can be used to obtain empirical estimates of the dependence matrix of the H\"usler--Reiss distribution.

Extensions that consider different tail decays have emerged in the last decade: for example,
\cite{beirlantetalIME11} and \cite{dtggoegui14} consider
projections of the form $Z_\omega=\min\{Y_1^{\mathrm{p}}, Y_2^{\mathrm{p}}\omega/(1-\omega)\}$ for $\omega \in (0,1)$ a fixed angle.
Under a regular variation assumption, the distribution of $Z_\omega$ can be approximated by the so-called extended Pareto distribution.
The parameters of the latter can be estimated using the minimum density power divergence (MDPD) criterion \citep{dtggoegui14}, which includes the maximum likelihood estimator as a special case.

The \pkg{RTDE} package \citep{RTDEpkg20} provides various functions to estimate the parameters of this model, and

the returned
 
objects allow users to summarize/plot fitted outputs, to compute the bivariate tail probability as well as to perform a simulation analysis. A similar approach is considered in \cite{Wadsworth:2013} and implemented in \code{lambdadep} function of the \pkg{mev} package; the authors

 look at different extrapolation paths by replacing the multivariate regular variation by a collection of univariate regular variation assumptions. \cite{Mhalla2019} also use such ideas to implement generalized additive regression for extremal dependence parameters.

 The drawback of these approaches, termed structural variables since they use univariate projections, is that estimation is carried independently for every angle $\omega$.

\subsection{Time series and graphical models} \label{sec:ts}
Data on a single variable collected over time often exhibit short-term temporal dependence, which can lead to extremes occurring in clusters. As a minimum, statistical methods for time series extremes need to account for dependence in the data and to estimate the extent to which extremes cluster, either directly or using a dependence model. For reviews of this area see \cite{Chavez-Demoulin/Davison:2012} and \cite{Reich/Shaby:2016}.

\subsubsection{Extremal index estimation} \label{sec:tsextremalindex}
For stationary processes satisfying the $D(u_n)$ condition, which limits long-range dependence at extreme levels, the strength of local serial extremal dependence is commonly measured by the extremal index. The latter can be interpreted as the reciprocal of the limiting mean cluster size in a Poisson cluster process of exceedances of increasingly high thresholds. \Cref{tab:eiestimators} gives basic information about the direct estimators of the extremal index that feature in this section, while \Cref{tab:eisoftware} summarises implementations of these estimators, including information about diagnostics for the choice of tuning parameters.  When a threshold is involved these diagnostics can be used for threshold selection. The diagnostics in the \pkg{evd}, \pkg{evir}, \pkg{exdex}, \pkg{fExtremes} and \pkg{texmex} packages are threshold stability plots for the extremal index. The information matrix test of \cite{Suveges/Davison:2010}, which is based on a model for truncated inter-exceedance times called $K$-gaps, is provided by the \pkg{mev} and \pkg{exdex} packages. The packages \pkg{evd} (function \code{clusters}), \pkg{extRemes} (\code{decluster}), \pkg{fExtremes} (\code{deCluster}), \pkg{POT} (\code{clust}) and \pkg{texmex} (\code{declust}) use an estimate of the extremal index to decluster exceedances of a threshold to form a series of sample cluster maxima.

\begin{table}
\centering
\begin{tabular}{lll}
\toprule
estimator & reference & tuning parameter(s) \\
\midrule
runs & \cite{Smith/Weissman:1994} & run length, threshold \\
blocks (blocks 1) & \cite{Smith/Weissman:1994} & block size, threshold \\
modified blocks (blocks 2) & \cite{Smith/Weissman:1994} & block size, threshold \\
intervals (FS) & \cite{Ferro/Segers:2003} & threshold \\
iterative least squares (ILS) & \cite{Suveges:2007} & threshold \\
$K$-gaps & \cite{Suveges/Davison:2010} & run length $K$, threshold \\
semiparametric maxima (SPM) & \cite{Northrop:2015} & block size \\
\bottomrule
\end{tabular}
\caption{Overview of some direct estimators of the extremal index with associated references and tuning parameters.}
\label{tab:eiestimators}
\end{table}

\begin{table}
\centering
\begin{tabular}{lllclll}
\toprule
package & estimator(s) & estimation & UQ & diagnostics  \\
\midrule
\pkg{evd} & runs, FS & \code{exi} & no & \code{exiplot}  \\
\pkg{evir} & blocks 2 & \code{exindex} & no & \code{exindex}  \\
\pkg{extRemes} & runs, FS & \code{extremalindex} & yes & ---  \\
\pkg{exdex} & ILS & \code{iwls} & no & ---  \\
  & $K$-gaps & \code{kgaps} & yes & \code{choose\_uk}  \\
   & SPM & \code{spm} & yes & \code{choose\_b}  \\
\pkg{fExtremes} & runs & \code{runTheta} & no & \code{exindexPlot}  \\
   & blocks 1 & \code{clusterTheta} & no & \code{exindexesPlot}  \\
   & blocks 2 & \code{blocktheta} & no &    \\
   & intervals & \code{ferrosegersTheta} & no &    \\
\pkg{mev} & ILS, FS & \code{ext.index} & no & \code{ext.index}  \\
   & $K$-gaps & \code{ext.index} & no & \code{infomat.test}, \code{ext.index}  \\
\pkg{POT} & runs & \code{fitexi} & no & \code{exiplot} \\
\pkg{revdbayes} & $K$-gaps & \code{kgaps\_post} & yes & ---  \\
\pkg{texmex} & FS & \code{extremalIndex} & yes & \code{extremalIndexRangeFit}  \\
\pkg{tsxtreme} & runs & \code{thetaruns} & yes & --- \\
\bottomrule
\end{tabular}
\caption{Comparison of \soft{R} packages for the direct estimation of the extremal index. Estimator(s): name(s) of the estimators available; estimation: function name(s) for estimation; uncertainty quantification (UQ): are methods for estimating uncertainty provided?; diagnostics: function names(s) for choosing tuning parameters.}
\label{tab:eisoftware}
\end{table}

\subsubsection{Marginal modeling} \label{sec:tsmarginal}
Suppose that interest is limited to marginal extremes.  The limiting distributions of cluster maxima and a randomly chosen threshold exceedance are identical, so inferences can be made using a marginal generalized Pareto model for sample cluster maxima or for all exceedances.  The \pkg{texmex} \citep{texmex} package is the most complete implementation of the analysis of cluster maxima: it uses a semi-parametric bootstrap procedure to account for uncertainty in declustering and in marginal inference and can also accommodate covariate effects. The declustering approach is wasteful of data and \cite{Fawcett/Walshaw:2012} show that the difficulty of identifying clusters reliably can lead to substantial bias.  When using all exceedances appropriate adjustment must be made for dependence in the data and for the value of the extremal index \citep{Fawcett/Walshaw:2012}: the \pkg{lite} package uses the methodology of \cite{Chandler/Bate:2007} to estimate a marginal log likelihood that has been adjusted for clustering using a sandwich estimator of the covariance matrix of the marginal parameters and combines this with a log likelihood for the extremal index under the $K$-gaps model.  The \pkg{extremefit} package provides a semiparametric procedure for time series extremes, as described in Section \ref{sec:nonpar}.  \Cref{tab:tssoftware} gives summaries of these packages and the packages that enable the estimation of time series dependence.

\subsubsection{Models for dependence} \label{sec:tsdependence}
In some applications it is important to infer more about the behavior of an extreme event than the size of a cluster of extreme values. For example, the duration of an extreme event or an accumulation of the extreme values may be of interest.  This requires the nature of serial extremal dependence to be modeled.   The \pkg{extremogram} package implements the extremogram \citep{Davis/Mikosch:2009, Davis/Mikosch/Cribben:2011, Davis/Mikosch/Cribben:2012} to inform modeling by exploring quantitatively serial extremal dependence within stationary time series and between different time series.  In the univariate case, it gives estimates of the conditional probabilities that a variable exceeds a user-supplied high threshold at time $t + l$ given that it exceeded this threshold at time $t$.  The stationary bootstrap is used to provide confidence intervals.

The \code{fitmcgpd} function in the \pkg{POT} package performs maximum likelihood inference using a first-order Markov chain model, in which one of several  bivariate extreme value distributions is used as a model for successive threshold exceedances \citep{Smith:1997}. The function \code{simmc} simulates from this type of model, as does the \code{evmc} function in the \pkg{evd} package.
The \pkg{tsxtreme} package models time series dependence using the conditional extremes approach of \cite{Heffernan:2004}, which enables a greater range of dependence structures to be modeled.  Inferences are performed using two-step maximum likelihood fitting and a Bayesian approach in which inferences are made about a more flexible model in which all inferences are performed simultaneously \citep{Lugrin/Tawn/Davison:2016}.  The functions \code{theta2fit} (\textsc{mle}) and \code{thetafit} (Bayesian) provide inferences for the sub-asymptotic extremal index of \cite{Ledford/Tawn:2003}.

The \pkg{ev.trawl} package implements the modeling approach described in \cite{Noven:2018}, which is based on the representation of a generalized Pareto distribution as a mixture of exponential distributions in which the exponential rate has a gamma distribution.  An exponential trawl process introduces time series dependence in a latent gamma process, while a marginal probability integral transform allows both negative and positive shape parameter values. The \pkg{CTRE} package deals with processes for which inter-exceedance times have a heavy-tailed distribution and therefore a Poisson cluster representation is not appropriate \citep{Hees2021}.  Parameter stability plots are provided as a means of selecting a suitable threshold.

\begin{table}
\centering
\begin{tabular}{lllclll}
\toprule
reference & package & function(s) & area  \\
\midrule
\cite{Fawcett/Walshaw:2012} & \pkg{texmex} & \code{declust}, \code{evm} & \textbf{m}  \\
\cite{Fawcett/Walshaw:2012} & \pkg{lite} & \code{flite} & \textbf{m} \\
\cite{Durrieu:2018aa} & \pkg{extremefit} & \code{hill.ts} & \textbf{m} \\
\cite{Davis/Mikosch:2009} & \pkg{extremogram} & \code{extremogram1}, \code{bootconf1}, $\ldots$ & \textbf{e} \\
\cite{Lugrin/Tawn/Davison:2016} & \pkg{tsxtreme} & \code{depfit}, \code{dep2fit} & \textbf{d} \\
\cite{Smith:1997} & \pkg{evd} & \code{evmc} & \textbf{d} \\
\cite{Smith:1997} & \pkg{POT} & \code{fitmcgpd, simmc} & \textbf{d} \\
\cite{Noven:2018} & \pkg{ev.trawl} & \code{FullPL, rtrawl} & \textbf{d} \\
\cite{Hees2021} & \pkg{CTRE} & \code{MLestimates} & \textbf{d} \\
\bottomrule
\end{tabular}
\caption{Overview of packages and main functions for modeling time series extremes by area: marginal modelling (\textbf{m}); exploratory analysis (\textbf{e}); dependence modelling (\textbf{d}).}
\label{tab:tssoftware}
\end{table}

\subsubsection{Graphical extremes}

Under the first-order Markov chain model for time series extremes of \cite{Smith:1997}, the value of a variable at time $t+1$ is assumed to be conditionally independent of its value prior to time $t$ given the value at time $t$. This simple dependence structure could be represented as a graphical model in which nodes representing the value of the variable are only connected by an edge if they correspond to adjacent time points.

The packages \pkg{graphicalExtremes} \citep{Engelke/Hitz:2020} and \pkg{gremes} \citep{Asenova:2021} provide more general graphical modeling frameworks for extremes, based on a multivariate H\"usler--Reiss generalized Pareto model for peaks over thresholds; see also \Cref{sec:mvthresh}. A graph represents conditional independences between variables.  If the graph is sparse then the joint distribution decomposes into the product of lower-dimensional distributions, which results in a more parsimonious and tractable model. If the graph is a tree, that is, there is exactly one path along edges between any pair of nodes, then this decomposition is particularly simple. The \pkg{graphicalExtremes} and \pkg{gremes} packages provide functions to fit a multivariate H\"usler--Reiss generalized Pareto model given a user-supplied graph and functions to simulate from this model.  The specifics of the theory underlying these packages differ but the resulting model structures coincide when based on a tree.

In some applications, such as the analysis of extreme river flows, there is a physical network from which the graph can be constructed.  In other cases the graph is conceptual: \pkg{graphicalExtremes} also provides a means to infer the structure of a graph from data.

\section{Functional extremes (including spatial extremes)}
\label{sec:functional}

\emph{Functional extremes} designates a relatively recent branch of extreme value analysis concerned with stochastic processes over infinite-dimensional spaces, especially  spatial and spatio-temporal extremes in  geographic space \citep{Davison2012,Huser2020}. We here use  the term \emph{space}  for $\mathbb{R}^d$ with $d\geq 1$, including the combination of geographic space and time ($d=3$),  and we explicitly refer to time  only where necessary. In practice, we usually work with finite discretizations of the study domain, such that many multivariate results and techniques carry over to the functional setting, although usually in relatively high dimension.

Common exploratory tools for extremal dependence are coefficients for bivariate distributions assessed as a function of spatial distance or temporal lag (e.g., extremal coefficient function based on bivariate extremal coefficients $\theta_2$, tail correlation function based on the $\chi$ measure, $F$-madogram, concurrence probability for maxima).

The asymptotic mechanisms for functional maxima and threshold exceedances  are similar to the multivariate setting. Available statistical implementations are summarized in \Cref{ssec:functional-extremes-summary-of-implementations}.  Marginal and dependence modeling is discussed in \Cref{ssec:functional-extremes-regression}.  Aspects that we consider as still underdeveloped in existing implementations are listed in \Cref{ssec:functional-extremes-what-is-missing}.

We use notation $\boldsymbol{X}=\{X(\boldsymbol{s})\}$ for stochastic processes  indexed by $\boldsymbol{s}\in \mathcal{S}\subset \mathbb{R}^d$, representing the process of the original event data. Usually we have  a dataset of observations $X_i(\boldsymbol{s}_j)$ for $j=1,\ldots, m$ locations observed at $i=1, \ldots, n$ time points. 

Max-stable processes are the natural class of models for locationwise maxima taken over temporal blocks of the same length, such as annual maxima observed at fixed spatial locations. A max-stable process possesses finite-dimensional max-stable distributions, and convergence to a max-stable process can be defined through the convergence of all finite-dimensional distributions, such that strong links arise with the univariate and multivariate setting.  If there exist sequences of normalizing functions $a_n(\boldsymbol{s})>0$ and $b_n(\boldsymbol{s})$ such that
\begin{align}\label{eq:conv-func-max}
a_n(\boldsymbol{s})\{M_n(\boldsymbol{s})-b_n(\boldsymbol{s})\}  \rightarrow Z(\boldsymbol{s}), \quad s \in \mathcal{S}, \quad n \to \infty,
\end{align}
with  $Z(\boldsymbol{s})$ a  nondegenerate limit process, then $Z(\boldsymbol{s})$ is max-stable.

Generalized $r$-Pareto processes  \citep{Ferreira2014,Thibaud2015,Dombry2015,deFondeville2018,Engelke2019} arise asymptotically  when a summary functional $r(\boldsymbol{X})$ of the process $\boldsymbol{X}$  exceeds a threshold that tends towards the upper endpoint of the probability distribution of the functional $r$.

Limit theory and statistical methodology  in this peaks over threshold setting  was formulated under the assumption that realizations of $\boldsymbol{X}$ correspond to continuous functions over a compact domain $\mathcal{S}$.
The most widely used setting for functional peaks over threshold follows the multivariate setting by assuming that data have been standardized, i.e., marginally transformed with a transformation $T$ that is strictly monotonic (i.e., $T(x_2)>T(x_1)$ if $x_2>x_1$), and that ensures positivity (i.e., $T(x)\geq 0$) with standardized tails of transformed random variables for which $\lim_{x \to \infty} x\Pr[T\{X(\boldsymbol{s})\}>x] =1$.
We can choose $T_{\boldsymbol{s}}\{X(\boldsymbol{s})\}=1/[1-F_{\boldsymbol{s}}\{X(\boldsymbol{s})\}]$, where $F_{\boldsymbol{s}}$ is the marginal distribution of $X(\boldsymbol{s})$.
Typically, summary functionals $r$ are homogeneous (i.e., $r(t\boldsymbol{x}) = tr(\boldsymbol{x})$ for $t>0$)
; examples include  the  average $r(\boldsymbol{x}) = |\mathcal{S}|^{-1}\int_{\mathcal{S}} x(\boldsymbol{s})\,\mathrm{d}s$, the minimum $r(\boldsymbol{x}) = \min_{\boldsymbol{s}\in\mathcal{S}} x(\boldsymbol{s})$, the maximum $r(\boldsymbol{x}) = \max_{\boldsymbol{s}\in\mathcal{S}} x(\boldsymbol{s})$, or  the median.  Convergence is assumed in the space of continuous functions over compact  $\mathcal{S}$, such that the distribution of the functional $r[T_{\boldsymbol{s}}\{X(\boldsymbol{s})\}] $ is well defined.
Functional convergence of maxima in \eqref{eq:conv-func-max} implies functional convergence to $r$-Pareto processes $\boldsymbol{Y}_r^\star= \{Y_r^\star(\boldsymbol{s})\}$:
\begin{align}\label{eq:conv-func-pot}
T_{\boldsymbol{s}}\{X(\boldsymbol{s})\}\mid r[T_{\boldsymbol{s}}\{X(\boldsymbol{s})\}] \geq u \to \{Y_r^\star(\boldsymbol{s})\} \quad u\to \infty.
\end{align}

Max-stable and generalized Pareto processes have different probabilistic structures, but there always is a one-to-one correspondence between their dependence structures.  Estimation of the marginal distributions and of the dependence structure is often conducted in two separate steps. The space-time dependence between sites is normally captured by correlation functions or variograms, which leads to much fewer parameters to infer than in the unstructured multivariate setting.

These asymptotic models can accommodate either asymptotic dependence or full independence among the variables $X(\boldsymbol{s}_1)$ and $X(\boldsymbol{s}_2)$ at locations $\boldsymbol{s}_1,\boldsymbol{s}_2\in\mathcal{S}$. The coefficient of tail dependence introduced in \eqref{eq:coef-tail-dep}, if considered for $D$ sites $\boldsymbol{s}_1,\ldots,\boldsymbol{s}_d$,  is therefore restricted to values $\eta\in\{1/D,1\}$. More flexible  dependence structures can be achieved within the conditional extremes framework with conditioning on a fixed location \citep{Wadsworth2019,Simpson2020}.  Finally, so-called subasymptotic models do not arise as classical extreme value limits but  focus on flexibly capturing dependence remaining at subasymptotic levels, for instance with asymptotic independence where $1/D<\eta(\boldsymbol{s}_1,\boldsymbol{s}_2)<1$ is possible; 

for example, the class of max-infinitely divisible processes \citep{Huser2021}, which is useful for flexible modeling of location-wise maxima. 
Most such proposals do not come with packaged and generic software implementations so far.

\subsection{Max-stable processes for maxima data}
\label{ssec:functional-extremes-summary-of-implementations}

Suppose that data consist of locationwise block maxima $M_i(s_j)$, where $i=1, \ldots, n_{\text{block}}$ indexes the blocks,  e.g., the observation year in case of annual block maxima. The \pkg{SpatialExtremes} package provides the most comprehensive collection of functions  for exploration and statistical inference with max-stable processes for spatial maxima data  in geographical space ($d=2$). While standard full likelihoods are not tractable even for moderately many locations with the common models, pairwise likelihood has become the standard approach for fitting max-stable processes, with implementations in \pkg{SpatialExtremes} and \pkg{CompRandFld}. Global dependence measures such as concurrence maps \citep{Dombry/Ribatet/Stoev:2018}, available from \code{concurrencemap} in \pkg{SpatialExtremes}, can be constructed from bivariate summaries. The intractability of the multivariate max-stable distribution function $G$, described in \cref{eq:bivariate:Glimit:1},  has led to pairwise likelihood becoming the standard estimation method for spatial max-stable processes. In \pkg{SpatialExtremes}, joint frequentist estimation of marginal and dependence parameters is possible, where auxiliary variables can be flexibly included in the three parameters of the marginal generalized extreme value distribution. Similar to generalized additive models, smoothness penalties can be imposed on nonlinear effects modeled through spline functions.  In contrast to the aforementioned generalized additive model approach without dependence, the numerical optimization becomes more involved here, such that only a moderate number of marginal parameters can be reasonably estimated.

\pkg{RandomFields} \citep{RandomFields} provides a large variety of max-stable models and particularly of tail correlation functions, with  a focus  on implementing simulation from such models. Moreover, the package  encapsulates vast functionality, especially simulation, for Gaussian random fields, which are often the building blocks for the more sophisticated extreme value models. The package provides multiple state-of-the-art algorithms for simulating Brown--Resnick max-stable processes. Exact unconditional simulation of max-stable processes is available in \pkg{RandomFields}, \pkg{mev} and \pkg{SpatialExtremes}, but only the latter offers conditional simulation of max-stable random fields (conditional on observed values at given locations) using Gibbs sampling \citep{Dombry/Eyi-Minko/Ribatet:2013}. \pkg{CompRandFld}'s simulation routine for max-stable processes uses an interface to \pkg{RandomFields}. For a particular hierarchically structured max-stable dependence model, known as the Reich--Shaby model \citep{Reich:2012} that is constructed using spatial kernel functions  and is derived from the spectral representation of a max-stable process based on a $l_p$-norm \citep{Oesting:2018}, estimation tools are available in the \pkg{hkevp} package and the experimental \code{abba} function in \pkg{extRemes}.  It is difficult to fit because of the dual role of its nugget parameter $\alpha>0$. The \pkg{hkevp} package provides a Metropolis-within-Gibbs algorithm for Bayesian estimation of the model and for simulation, whereas the \code{abba} is flagged as experimental and is less comprehensive.

\subsection{Peaks-over-threshold modeling}
\label{ssec:functional-extremes-ell-Pareto-processes}

For functional peaks over threshold, \pkg{mvPot} \citep{mvPot} provides parametric simulation and estimation tools for various $r$-Pareto processes using Brown--Resnick and extremal Student-$t$ dependence structures. Parameter estimates are computed using optimization of either full likelihood or gradient score functions; the latter remains computationally tractable for settings where full likelihood does not. Estimation of the marginal transformation $T$ is not implemented  and has to be performed prior  to estimating the extremal dependence parameters using \pkg{mvPot}. A competitive estimation procedure is the gradient score estimating equation of \cite{deFondeville2018}, which does not require calculation of the normalizing constant of the model and also replaces censoring with downweighting. While statistically less efficient than full likelihood estimation, the procedure is more robust and can be applied in very high-dimensional settings. For estimation, numerical implementations are currently restricted to the Brown--Resnick model  but tools for simulation and calculation of likelihoods are  available for the extremal-$t$ dependence model. The \pkg{mev} package also proposes likelihood functions and unconditional simulation routines for generalized $r$-Pareto processes \citep{deFondeville:2021}.

Some other implementations allowing estimation of asymptotic dependence structures use  original event data $X_i(s_j)$ and can be viewed as working on the interface of max-stable and peaks over threshold models. For example, moment-based estimation of parametric models, based on contrasting empirical and parametric versions of a variant of the so-called tail dependence function, is implemented in the package \pkg{tailDepfun} \citep{Einmahl/Segers/Kiriliouk:2018}.

\subsection{Modeling spatially varying marginal distributions}
\label{ssec:functional-extremes-regression}

In practice, marginal distributions $F_{\boldsymbol{s}}$  in functional data are usually not stationary, such that variation of marginal extreme value parameters with respect to space and time, or with respect to other available auxiliary variables, has to be captured. In the locationwise maxima setting, we can use use the generalized extreme value distribution and consider its parameters as functions of space, i.e., $\xi(\boldsymbol{s}),\mu(\boldsymbol{s}),\sigma(\boldsymbol{s})$. Different options exist in the peaks over threshold setting. A common approach is to fix a high, potentially nonstationary threshold $u(\boldsymbol{s})$, and then estimate the threshold exceedance probability $p(\boldsymbol{s})=\Pr\{X(\boldsymbol{s})>u(\boldsymbol{s})\}=1-F_{\boldsymbol{s}}\{u(\boldsymbol{s})\}$ and the generalized Pareto parameters $\xi(\boldsymbol{s}),\sigma(\boldsymbol{s})$ based on observations of the exceedances $X(\boldsymbol{s})-u(\boldsymbol{s})>0$.

The regression frameworks discussed in Section~\ref{ssec:functional-extremes-regression} are relevant for modeling marginal extreme value parameters that vary with location in a first modeling step. Generalized additive modeling allows capturing complex nonlinear patterns of spatial nonstationarity  using relatively large numbers of parameters. Some care may be required in tuning smoothing hyperparameters since in this step one usually assumes independence of observations $X_i(s_j)$, so functional  dependence across space or time is disregarded. Specifically, MCMC-based Bayesian estimation of marginal parameters (using Gaussian process priors) for generalized extreme value distributions for maxima is possible through \pkg{SpatialExtremes} \citep{SpatialExtremes}, and \pkg{hkevp} \citep{hkevp} includes a similar function. The \pkg{SpatialGEV} package \citep{Chen/Ramezan/Lysy:2021} provides a template for fitting latent spatial models with marginal generalized extreme value distributions and Gaussian process priors on the parameters using quadratic approximations to the marginal posterior. The unpublished package \href{https://github.com/NorskRegnesentral/SpatGEVBMA/}{\pkg{SpatGEVBMA}}, formerly on the CRAN, fits a latent model with generalized extreme value margins whose parameters follow Gaussian process priors with explanatory variables. Its defining functionalities are the use of Laplace approximation for automating proposals and efficient exploration of the posterior, and  Bayesian model averaging to account for variable selection uncertainty \citep{Dyrrdal:2014}.

An important alternative to Monte Carlo methods is to estimate complex integrals arising from \Cref{eqbayes} through the integrated nested Laplace approximation (INLA). The \pkg{INLA}  package proposes computationally convenient representations of the spatial Mat\'ern covariance function through  the stochastic partial differential equation approach of \citet{Lindgren2011} for spatial and spatio-temporal latent Gaussian modeling. As mentioned in \Cref{sec:bayesian}, \pkg{INLA} provides implementation for generalized extreme value distributions (with covariates and random effects in the location parameter) and the generalized Pareto distribution (with covariates and random effects in a quantile at a probability level $\alpha\in(0,1)$ specified by the user; see \citet{Opitz2018} and \citet[][Chapter~6]{Krainski2018}.

The package further allows joint estimation of several regression designs where some of the random effects can be in common (i.e., shared through a scaling coefficient) among these, which is beneficial to obtain cross-correlation in the posteriors of the predictors of several response types. For example, we could combine a logistic regression for the exceedance probability with a generalized Pareto regression for the excess above the threshold, and a shared random effect with a positive sharing coefficient would entail positive posterior correlation between the exceedance probability and the size of the excess.

\subsection{Outlook for functional extremes}
\label{ssec:functional-extremes-what-is-missing}

The coverage of max-stable processes, which remains an area of very active research, is much more comprehensive than others, with the notable exception of composite or full log likelihood inference for max-stable processes. Formulae exist for many partial derivatives of the exponent function $V$ arising in the multivariate max-stable cumulative distribution functions and, in principle, the  Stephenson--Tawn likelihood (or a  bias-corrected version thereof) could be programmed for full likelihood inference beyond the bivariate case. Most of the models are also implemented with spatial applications in mind, even if temporal or spatio-temporal applications are possible. Max-infinitely divisible models are not covered in software yet, and Bayesian models with latent processes are often not provided with numerical implementation because of the complexity of implementation and also sometimes very long execution times of codes.

There are much fewer implementations for threshold models. Whereas their construction can be viewed as more flexible and intuitive than the one of the corresponding max-stable processes, they are conditional models with respect to threshold exceedance of the summary functional $r$. In the finite-sample setting of statistical practice, this means that  observations at some locations may not correspond to marginal exceedances and may therefore not  be coherent with the asymptotic model. A common remedy is censoring, but this makes estimation more costly because the likelihood functions now include high-dimensional distribution functions which typically must be calculated via Monte-Carlo methods for each vector of observation. Generic full likelihood estimation procedures have been proposed, and are implemented  (though computationally costly) for some models. However, available implementations do not yet come with a comprehensive set of available models and methods for parameter inference, model validation and comparison. An obvious solution to facilitate such implementation, provided that parameters are identifiable from lower-dimensional summaries, would be to use composite likelihood, but this is not implemented. Likewise, Bayesian generalized Pareto models with latent Gaussian process priors could be easily implemented in many probabilistic programming languages outside of \soft{R}, such as \soft{Stan}, but no general-purpose routines exist so far.

Simulation algorithms for unconditional simulation from generalized $r$-Pareto processes with arbitrary risk functionals $r$ are still elusive, as designing efficient accept-reject methods requires case-by-case analysis. Available conditional simulation code typically amounts to simulation of elliptical distributions (log-Gaussian or Student-$t$) with linear constraints.

Implementations with documented code are often available as supplementary material to methodological papers but have not been encapsulated in officially validated packages; see \citet{Huser2019,Bacro2020,Simpson2020} for recent examples. \citet{Huser2019} has companion code for frequentist estimation of a flexible subasymptotic spatial model in the unpublished package \pkg{spatialADAI}.  An INLA-based implementation for Bayesian conditional extremes models for spatial and spatio-temporal data is provided for \cite{Simpson2020} as supplementary material. The implementation of many Bayesian extreme value models in the literature is achieved with standard MCMC algorithms that are tailored to the particular data application, but often generic and easily reusable or reproducible code is not provided, which hinders reproducibility.

  \begin{table}
\centering
\begin{tabular}{lllll}
\toprule
methods & package & functions & scope\\ \midrule
& \pkg{copula} & \code{rCopula}${}^*$ & \textbf{b}, \textbf{m} \\
 \cite{Tajvidi/Turlach:2018} & \pkg{SimCop} &   --- &\textbf{b}\\
   \cite{Stephenson:2003} & \pkg{evd} & \code{rbvevd}, \code{rmvevd} & \textbf{b}, \textbf{m} \\
   \cite{Dombry:2016} & \pkg{mev} & \code{rmev}, \code{rmevspec}  & \textbf{a}, \textbf{m}, \textbf{f}\\
  \cite{Engelke/Hitz:2020} & \pkg{graphicalExtremes} & \code{rmstable} & \textbf{m}, \textbf{f} \\
  \cite{Beranger:2017} & \pkg{ExtremalDep} & \code{r\_extr\_mod} & \textbf{m}, \textbf{f}\\
  \cite{CompRandFld} & \pkg{CompRandFld} & \code{RFsim}${}^*$ & \textbf{f}\\
    \cite{Dombry/Eyi-Minko/Ribatet:2013}& \pkg{SpatialExtremes} & \code{condrmaxstab} & \textbf{f} \\
    \cite{Dombry:2016} & \pkg{SpatialExtremes} & \code{rmaxstab} & \textbf{f} \\
 \cite{RandomFields}   & \pkg{RandomFields} & \code{RFsimulate}${}^*$ & \textbf{f}\\
  \cite{Reich:2012} & \pkg{hkevp} & \code{hkevp.rand} & \textbf{f}\\
  \cite{Ballani:2011} & \pkg{BMAmevt} & \code{rnestlog}, \code{rpairbeta} & \textbf{a}\\
\cite{Ho:2019} &\pkg{mev} & \code{rparpcs} & \textbf{p}\\
\cite{deFondeville2018} & \pkg{mev} & \code{rparp} & \textbf{p}\\
\cite{deFondeville2018} & \pkg{mvPot} & --- & \textbf{p}\\
\cite{deFondeville:2021} &\pkg{mev} & \code{rgparp} & \textbf{p} \\
 \bottomrule
\end{tabular}
\caption{Overview of simulation algorithms for bivariate (\textbf{b}) and multivariate (\textbf{m}) max-stable distributions  and for max-stable processes (\textbf{f}), and for angular (\textbf{a}) and Pareto processes (\code{p}) with associated references. Some of the listed functions (${}^*$) are generic and include specific classes for max-stable models, but other models as well.}
\label{tab:simulation}
\end{table}

  \begin{table}
\centering
\scalebox{0.92}{
\begin{tabular}{sllll}
\toprule
reference & package & functions & dim & data \\
\midrule
\cite{Coles:1991} & \pkg{evd} & \code{fbvevd} & \textbf{b} & max \\
 & \pkg{copula} & \code{fitCopula} & \textbf{b} & max\\
 \cite{Einmahl/Segers/Kiriliouk:2018} &\pkg{tailDepFun} & \code{Estimation}$\cdots$ & \textbf{m}, \textbf{f} & max\\
\cite{pickands81} and  &\pkg{evd} & \code{abvnonpar} & \textbf{b} & ang\\
\quad \cite{caperaafougeregenest97} & & &  \\
\cite{Gudendorf/Segers:2012} &\pkg{copula} & \code{An} & \textbf{b}, \textbf{m} & ang\\
 \cite{Einmahl/Segers:2009} and& \pkg{extremis} & \code{angcdf} & \textbf{b} & ang\\
 \quad \cite{deCarvalho:2013}& \pkg{mev} & \code{angmeas} & \textbf{m} & ang\\
\cite{Marcon:2017} & \pkg{ExtremalDep} & \code{madogram}, \code{beet} & \textbf{m} & ang\\
\cite{Wadsworth2016} & \pkg{EVcopula} & \code{fit.EV.copula} & \textbf{b} & ang\\
\cite{Sabourin:2014} & \pkg{BMAmevt} & \code{posteriorMCMC} & \textbf{m} & ang \\
\cite{Smith:1997} & \pkg{evd} & \code{fbvpot} & \textbf{b} & pot \\
\cite{Engelke2019} & \pkg{graphicalExtremes} & \code{fmpareto\_graph\_HR} & \textbf{m} & pot \\
\cite{Heffernan:2004} & \pkg{texmex} & \code{mex} & \textbf{m} & pot\\

\cite{Davison2012} &\pkg{SpatialExtremes} & \code{fitcopula}, \code{fitmaxstab} & \textbf{f} & max \\
\cite{CompRandFld} & \pkg{CompRandFld} & \code{FitComposite} & \textbf{f} & max\\
\cite{Reich:2012} & \pkg{hkevp} & \code{hkevp.fit} & \textbf{f} & max\\
\cite{Reich:2012} & \pkg{extRemes} & \code{abba} & \textbf{f} & max\\
\bottomrule
\end{tabular}
}
\caption{Overview of multivariate and functional estimation procedures for extremes based on dimension, either  bivariate (\textbf{b}), multivariate (\textbf{m}) or functional (\textbf{f}) and data type/paradigm, one of block maximum (max), pseudo-angles (ang) or threshold exceedances (pot). Packages which only include likelihood but no wrapper are excluded.}
\label{tab:fitmulti}
\end{table}
\section{Specialized topics}

While our review has ranged mostly over software providing implementation of relatively generic methods that can be useful in various application contexts, there also has been active development of software libraries targeting specific application fields, and we here cite some of them.

\noindent \textit{Hydrology and climate}: Regional frequency analysis using $L$-moments is possible with the \pkg{lmomrfa} package. The \pkg{climextRemes} package  leverages \pkg{extRemes} for climate extremes and implements methods highly relevant for this field, such as local likelihood fitting; the package is also available in \soft{Python}. \pkg{IDF} provides intensity-duration-frequency (IDF) curves \citep{Ulrich:2020}. \pkg{jointPm} implements the method of \cite{Zheng/Leonard/Westra:2015} for evaluating bivariate probabilities of exceedance.  An example of a highly specialized package is \pkg{futureheatwaves} and facilitates finding, characterizing and exploring heatwaves in climate projections, while the \soft{Python} package \pkg{teca} is dedicated to tracking extremes of large scale climate models. \pkg{Renext} includes methods for peaks over threshold with a variety of distributions and the possibility to include historical maximum records, along with tests of exponentiality and goodness-of-fit.

\noindent \textit{Financial and actuarial science}: Some packages provide implementation of various  generic models and methods for extreme values, but  make strong use the semantics of those fields in their documentation.
The packages \pkg{QRM} (and its successor \pkg{qrmtools}) and \pkg{ReIns} implement various functions to accompany the books \cite{mcneil2015quantitative} and \cite{albrecher2017reinsurance}, respectively.

The package \pkg{fExtremes} \citep{fExtremes} provides functions for financial analysis used by the \href{https://www.rmetrics.org/}{\textbf{Rmetrics}} project.

The package \pkg{VaRES} \citep{VaRES} provides two popular risk measures (value at risk and expected shortfall) for a large collection of probability distributions, including many heavy-tailed distributions. The \pkg {extremis} package proposes functionalities to cluster multivariate financial time series based on their frequency and magnitude of extreme events.

\noindent \textit{Machine learning}: The interface between statistical machine learning and extreme values has been growing in recent years, with proposals encompassing the use of gradient boosting for extremes (\citet{Velthoen:2021}, unpublished package \href{https://github.com/JVelthoen/gbex/}{\pkg{gbex}}) and of extremal random forests (\citet{Gnecco:2022}, package \pkg{erf}) for modeling high quantiles of a univariate distribution. Another area of active research is open-set classification, dealing with classification of observations in categories not observed in training data: the \soft{Python} package \pkg{EVM} implements the extreme value machine of \cite{Rudd:2018}, whereas \soft{R} package \pkg{evtclass} includes the algorithms described in \cite{Vignotto/Engelke:2020}.

\noindent \textit{Survival analysis}: Presence of censoring or truncation mechanisms, common in survival analysis, require dedicated software implementations because they affect the likelihood contribution of observations. The \soft{Matlab} \pkg{LATools} \citep{Rootzen/Zholud:2017} proposes an interface for interval-truncated generalized Pareto observations, while the \pkg{longevity} package handles more general partial observation schemes.

\section{Discussion and conclusion}

We have covered in this review a wide range of available software implementations for extremes, and we sincerely hope without omissions that are considered as important by authors or users.
The development of extreme value software is key to  extreme value analysis in practice and has become an active area of research,  but the availability of implementations tends to lag behind methodological innovations since these are often not accompanied by generic,  easily reusable and validated codes.
To encourage modelers in applied sciences and in operational services to make use of the most advanced methods and models, off-the-shelf implementations  are desirable. However, generic code  may be difficult to provide due to the high sophistication of approaches as, for instance, with functional extremes. Designing generic estimation procedures that are flexible enough to be useful while at the same time being robust requires particular care. Writing this review made us aware of how challenging it is for the extreme value community to develop tested and easily reusable software that  keeps pace with methodological progress: most software was written a decade ago, there are only handful of active maintainers, and most models proposed in the literature are not put together with software.

Many methods proposed in the last two decades are still not available; lack of implementation of bespoke code is a major impediment for their adoption. The most obvious gap is in software for fitting multivariate max-stable models (with composite likelihood) and multivariate generalized Pareto distributions with censoring in moderate dimensions for the parametric models with suitable tools. The conditional spatial extremes model, which extends the Heffernan--Tawn approach to the spatial setting, has been used in many recent papers but no software has been released.

More methods and more refined tools are also required for the nonstationary exploration and inference of extreme values. In many application fields, physical change processes (e.g., climate change, land-use change) require tools to explore, model and infer nonstationary behavior in extremes, for instance for climate-change detection and attribution. Currently, nonstationary modeling is implemented for marginal distributions through regression designs, but implementations providing dedicated methods for extreme value detection and attribution under climate change are scarce, and easily reusable codes  for nonstationary extreme value dependence structures are yet missing.

\section*{Acknowledgements}
Funding in partial support of this work was provided by the Natural Sciences and Engineering Research Council (RGPIN-2022-05001, DGECR-2022-00461).

\bibliographystyle{apalike2}
\phantomsection 
\addcontentsline{toc}{section}{Bibliography} 

\nocite{mev,ercv,Renext}
\nocite{exdex}
\nocite{lax}
\nocite{lite}
\nocite{revdbayes}
\nocite{threshr}
\bibliography{extreme,rpackages}

\end{document}